\documentclass[acmtog]{acmart}

\usepackage{amsmath}
\usepackage{pifont}
\usepackage{textcomp}
\usepackage{booktabs} 
\usepackage{microtype}
\usepackage{bm}
\usepackage{overpic}
\usepackage{siunitx}
\usepackage{wrapfig}
\usepackage[utf8]{inputenc}
\usepackage{multirow}
\usepackage{colortbl}
\usepackage{nicefrac}
\usepackage{xr}
\usepackage{enumerate}
\usepackage{hyphenat}
\usepackage[shortcuts]{extdash}
\usepackage[ruled]{algorithm2e} 

\SetAlFnt{\small}
\SetAlCapFnt{\small}
\SetAlCapNameFnt{\small}
\SetAlCapHSkip{0pt}

\acmJournal{TOG}

\definecolor{figred}{rgb}{1,0,0}
\definecolor{figgreen}{rgb}{0,0.6,0}
\definecolor{figblue}{rgb}{0,0,1}
\definecolor{figpink}{rgb}{1,0.63,0.63}
\definecolor{BLUE}{rgb}{0,0,1}

\newcommand{\figref}[1]{Fig.~\ref{fig:#1}}

\newcommand{\secref}[1]{\S\ref{sec:#1}}

\newlength\savedwidth

\graphicspath{{figs/}}

\begin{document}
\title{Principles towards Real-Time Simulation of Material Point Method on Modern GPUs}

\author{Yun (Raymond) Fei}
\affiliation{%
 \institution{Tencent Game AI Research Center}
 \city{Los Angeles}
 \country{USA}}
\email{raymondfei@tencent.com}
\author{Yuhan Huang}
\affiliation{%
 \institution{NVIDIA}
 \city{Shenzhen}
 \country{China}}
\email{akanh@nvidia.com}
\author{Ming Gao}
\affiliation{%
 \institution{Tencent Game AI Research Center}
 \city{Los Angeles}
 \country{USA}}
\email{mingwiscgao@tencent.com}


\begin{abstract}
Physics-based simulation has been actively employed in generating offline visual effects in the film and animation industry.
However, the computations required for high-quality scenarios are generally immense, deterring its adoption in real-time applications, e.g., virtual production, avatar live-streaming, and cloud gaming.
We summarize the principles that can accelerate the computation pipeline on single-GPU and multi-GPU platforms through extensive investigation and comprehension of modern GPU architecture. We further demonstrate the effectiveness of these principles by applying them to the material point method to build up our framework, which achieves $1.7\times$--$8.6\times$ speedup on a single GPU and $2.5\times$--$14.8\times$ on four GPUs compared to the state-of-the-art.
Our pipeline is specifically designed for real-time applications (i.e., scenarios with small to medium particles) and achieves significant multi-GPU efficiency. We demonstrate our pipeline by simulating a snow scenario with 1.33M particles and a fountain scenario with 143K particles in real-time (on average, 68.5 and 55.9 frame-per-second, respectively) on four NVIDIA Tesla V100 GPUs interconnected with NVLinks.
\end{abstract}

%
%

%

%
%


\maketitle
\setlength{\abovedisplayskip}{0.5ex}
\setlength{\belowdisplayskip}{0.5ex}
\setlength{\abovedisplayshortskip}{0.5ex}
\setlength{\belowdisplayshortskip}{0.5ex}

\begin{figure*}[t]
\centering
    \includegraphics[width=1.0\linewidth]{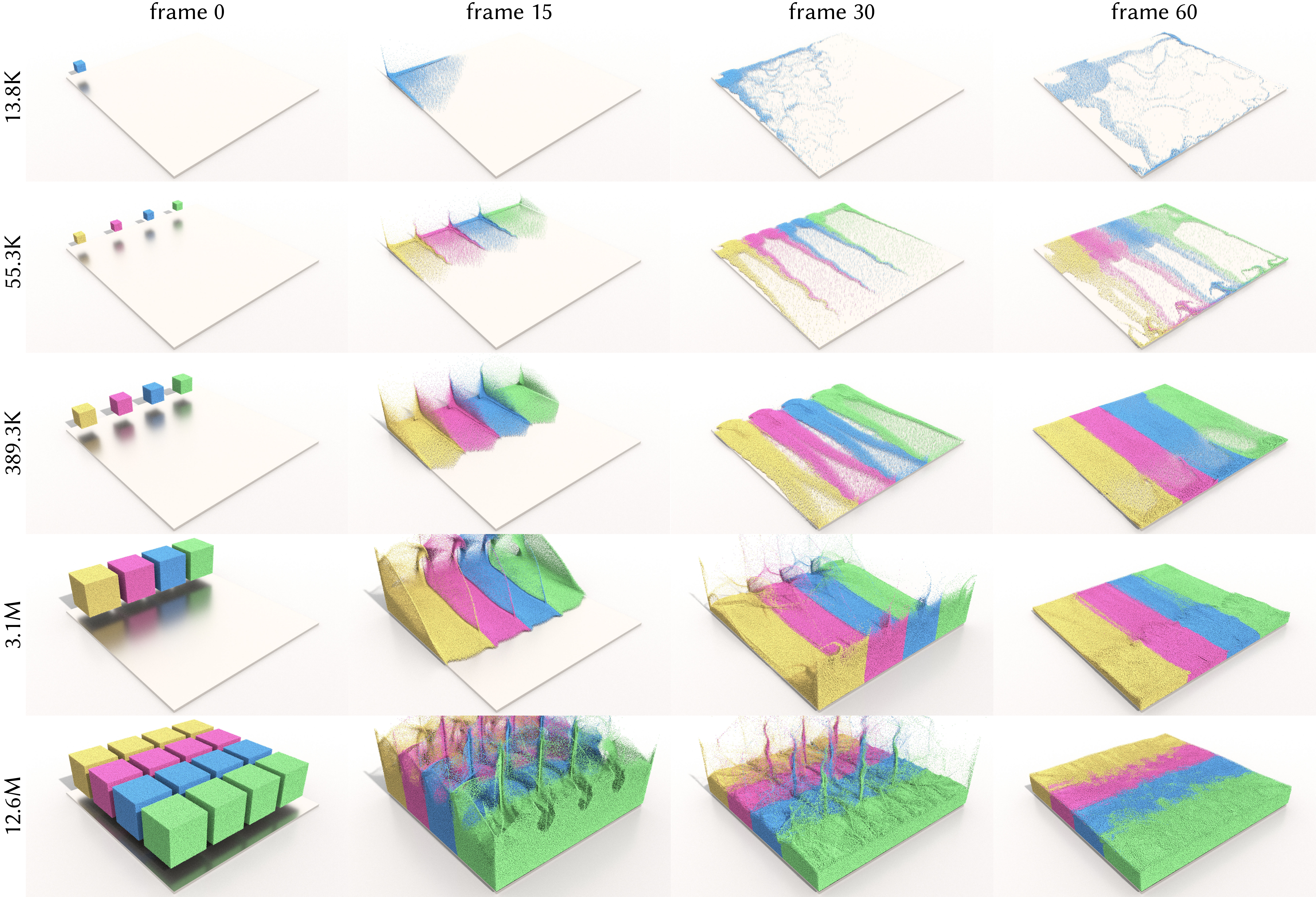}
    \includegraphics[width=1.0\linewidth]{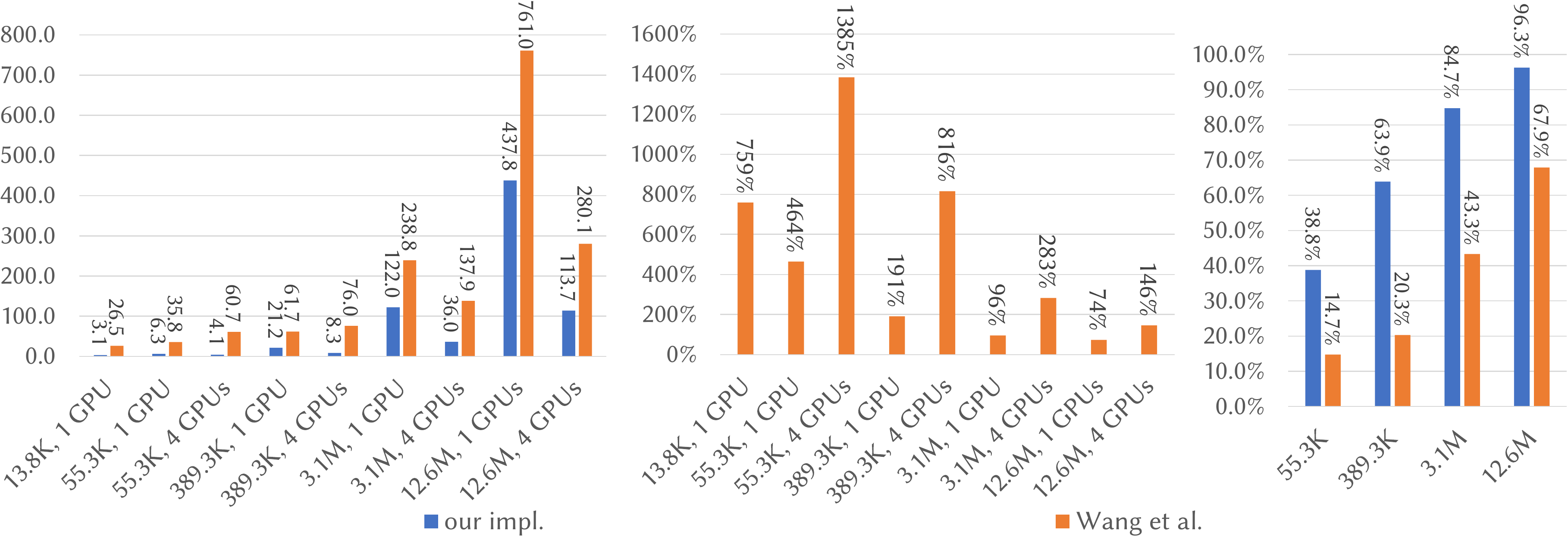}
    \caption{\textbf{Top: \emph{Sand Blocks.}} Blocks of sand with different sizes are dropped into a static, frictionless box, where particles simulated with different GPUs are colored differently. \textbf{Bottom (from left to right):} Timing per frame (in milliseconds, averaged over 60 frames) comparison between our implementation and Wang et al.~\shortcite{wang2020massively}'s, the latter's percentage of \emph{additional costs} per frame compared with ours, and the comparison of four-GPUs' efficiencies computed through equation (\ref{eq:efficiency}) from the averaged timings.}
    \label{fig:compare_wang}
\end{figure*}
\section{Introduction}
Cinematic narratives in modern games and other real-time applications (e.g., virtual production, avatar live-streaming, and cloud gaming) are becoming increasingly extensive, causing a great demand for more refined visual effects production. Creating an atmosphere that meets the plots' requirements for natural scenery may require realistic snow, sand, and water simulations.

On the other hand, using physical simulation to generate large-scale, plausible natural or character-related effects has become a standard in the production of films. Material point method (MPM)~\cite{sulsky1994particle} is widely used due to its quantitative accuracy~\cite{gaume2018dynamic} and strong generalization capability, enabling the plausible simulation of various complex materials and the coupling between them~\cite{jiang2016material,hu2019hybrid}. However, the computations required for high-quality scenarios are generally immense, deterring its adoption in real-time applications. Recently, to make the MPM more efficient, Gao et al.~\shortcite{gao2018gpu} developed MPM on a graphics processing unit (GPU), and Wang et al.~\shortcite{wang2020massively} developed MPM on multiple GPUs. At the same time, Hu et al.~\shortcite{hu2019taichi} also revealed the Taichi framework, which is more convenient for developing MPM on the GPU. 

Compared to MPM implementations on the CPU, these methods or frameworks have significantly improved the performance. However, there is still a gap between the achieved performance and the performance required by real-time applications. Besides, these works mainly focus on optimizing performance in the case of large numbers of particles. Few works have been done to minimize the extra cost of memory access or synchronization, which are critical for real-time applications that usually require a small number of particles but with a much tighter temporal budget.

In order to further improve the performance of GPU-MPM to reach the requirements of real-time applications, we provide four contributions:
\begin{enumerate}
\item optimization strategies that significantly improve the performance of MPM on single or multiple GPUs, providing excellent scaling on a multi-GPU platform even when the number of particles is relatively small (e.g., less than 100K);
\item ablation tests of various performance optimization strategies on GPU, proposed in either this article or previous works. Besides, we provide analysis in the aspects of hardware architecture and GPU instruction sets;
\item choices of different strategies based on these test and analysis results, and a more efficient implementation for simulations at different scales, which achieves $1.7\times$--$8.6\times$ speedup on single-GPU and $2.5\times$--$14.8\times$ acceleration on four GPUs compared with the state-of-the-art~\cite{wang2020massively}; and
\item principles that summarize our choices to inspire further optimization of the performance of MPM and other similar methods.
\end{enumerate}
\section{Related Works}

\subsection{Material Point Method}
The material point method (MPM) is a comprehensive simulation framework proposed by Sulsky et al.~\shortcite{sulsky1994particle}. MPM uses both Lagrangian and Eulerian representations and has been used in many works in mechanics~\cite{ZHANG201792} as well as computer graphics~\cite{jiang2016material,hu2019hybrid} to simulate complex phenomena including fluids, solids, and coupling between different materials. In this work, we adopt the material point method and focus on its code optimization on the GPU. Our optimization principles apply to MPM and also can be helpful to other similar frameworks using a mixed Lagrangian-Eulerian representation (e.g., fluid simulation with an implicit pressure solver~\cite{bridson2015fluid}).

\subsection{GPU Based Real-Time Physical Simulation}
With the rapid development of computing architectures and software stacks, GPU has become a powerful tool in real-time physical-based simulation. GPU has the benefits of better energy efficiency and higher per-dollar performance. 
However, migrating a CPU simulation algorithm to the GPU without significant performance degradation is nontrivial. The GPU has SIMD architecture and unique memory models that require special treatment on memory access and program design. 

\textit{Single GPU Simulation}.
Most research works focus on acceleration on a single GPU. Recently, there are research works using a single GPU to accelerate MPM solvers~\cite{gao2018gpu}, Finite Element Methods (FEM) solvers~\cite{Bernstein2016,wang2016descent}, Eulerian fluids solvers~\cite{chentanez2011real, cohen2010interactive}, pure-Lagrangian fluid solvers~\cite{goswami2017real, amada2004particle, Macklin2014, chen2015wetbrush, winchenbach2016constrained}, thin-film solvers~\cite{Vantzos2018gpu}, cloth solvers~\cite{tang2016cama,tang2018cloth,wu2020safe,wang2021gpu}, and other hybrid Eulerian-Lagrangian solvers~\cite{Chentanez2015,Wu2018}. Additionally, Hu et al.~\shortcite{hu2019taichi} proposed a framework where the user may write efficient simulation code on the GPU more easily.

\textit{Multiple GPU Simulation}.
A common expectation is that multiple GPUs can provide higher computational power and performance than a single GPU. However, physical simulation with multiple GPUs is more challenging than a single one due to the limited communication bandwidth. Besides, the overhead of launching a CUDA API (e.g., \emph{cudaMallocAsync}) increases along with the number of CPU-GPU pairs due to the required sophisticated locking mechanism used in these APIs~\cite{abe2014gdev}.
The overhead brought by multiple GPUs becomes even more significant when pursuing real-time simulation. The computational scale is usually tiny for real-time cases, and most time can be spent on communication if the code is not well optimized. 
Among the recent research, Herman et al.~\shortcite{Hermann2010} introduced a cooperative framework for interactive physical simulation using both multi-GPU and multi-CPU.
They use a multiple-GPU abstract layer to handle memory transfer between different GPUs and handle job splitting dynamically.
Junior et al.~\shortcite{Junior2012} implemented an architecture for simulation of real-time fluid using the multi-GPU SPH method.
They used a domain splitting method to distribute jobs to multiple GPUs and approached a $1.19\times$--$1.8\times$ speed up on two GPUs.
For MPM, Wang et al.~\shortcite{wang2020massively} implemented a multi-GPU MPM, which mainly focuses on simulations with tens of millions of particles. More recently, Li et al.~\shortcite{li2020p} presented a multi-GPU cloth solver, which focuses on simulations with millions of triangles.

Most prior works focus on accelerating MPM simulation performed in a large-scale, off-line environment. This paper shows that a real-time MPM simulation may have different considerations, and we attempt to push multi-GPU acceleration to the real-time domain. We proposed several principles, with which a real-time four-GPU MPM simulation with 55K particles can be $14.8\times$ faster than the state-of-the-art, and $2.5\times$ faster for off-line simulation with 12M particles.

\section{Principles on Modern GPU\label{sec:principles}}
This section summarizes and briefly discusses the principles and leaves the elaborated study to later sections.
For real-time high-quality physics-based simulation, we would like to assume that the graphics cards we used to have a large GPU memory size with NVLink connectivity~\cite{foley2017ultra} between each pair of them.
While all principles are proposed for designing a highly efficient simulator (e.g., applying them to an incompressible fluid solver), in this work, we mainly focus on optimizing the pipeline of the \emph{explicit material point method} to demonstrate their efficacy.

\subsection{Single GPU}
We summarize the principles for the computation on a single GPU as follows,
\begin{itemize}
    \item Avoiding intrinsic functions without native hardware support.
    \item Reducing memory reallocation once the simulation starts;
    \item Minimizing the number of CUDA kernels executed within a single time step;
    \item Minimizing the synchronization between GPU and CPU;
    \item Fine-tuning the CUDA block size and the usage of on-chip memory.
\end{itemize}
The first three principles strongly impact the performance, from which the schemes derived contribute more than half of the performance improvement. A detailed ablation study is presented in~\secref{ablation_single}.

\begin{figure}[htb]
\centering
    \includegraphics[width=1.0\linewidth]{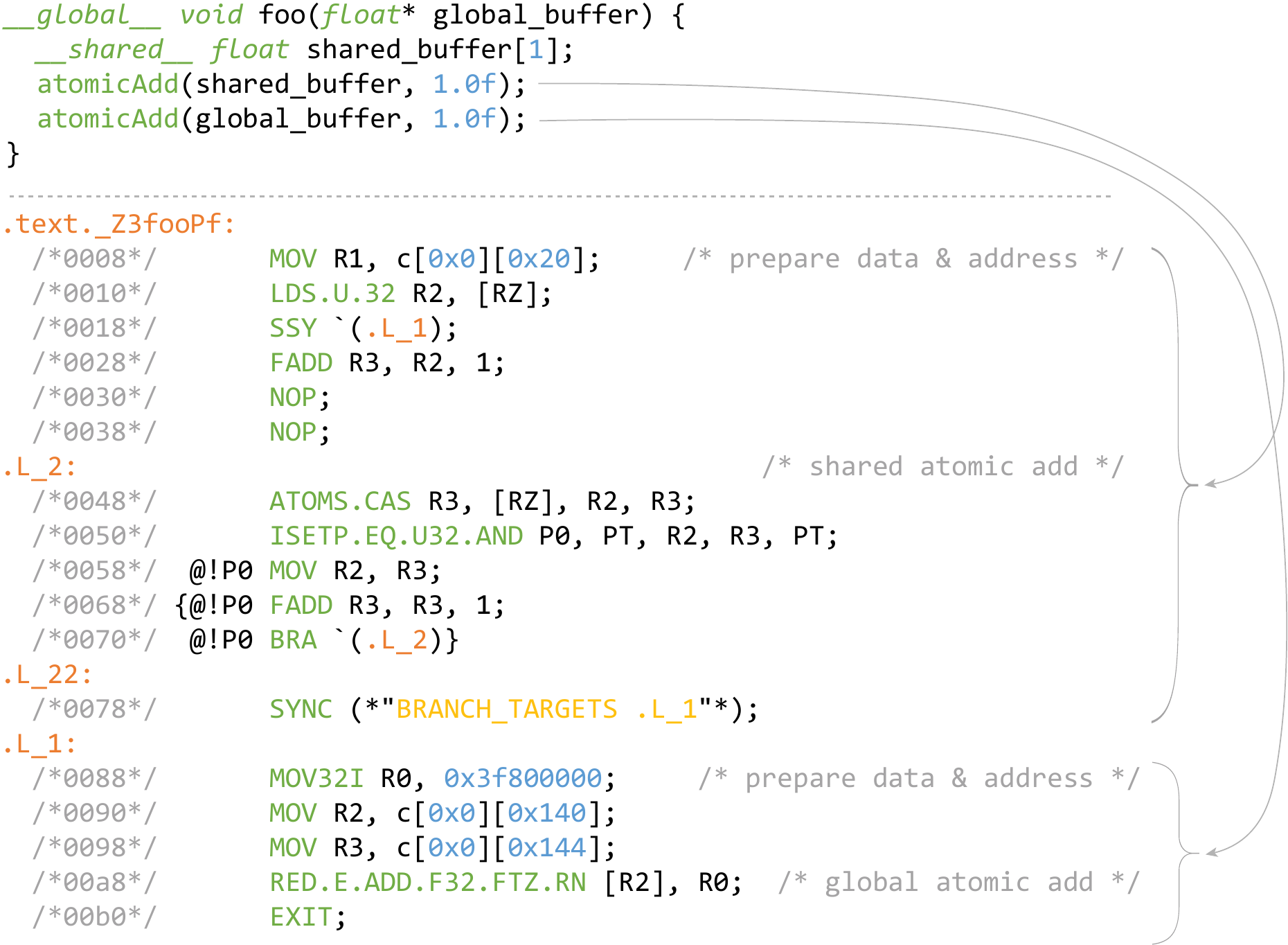}
    \caption{\textbf{Top:} A didactic kernel function adding a scalar atomically into shared and global memory; \textbf{Bottom:} The \texttt{SM\_70} disassembly (or \textbf{S}treaming \textbf{ASS}embly, \textbf{SASS}) of the compiled binary code. The (non-native) shared atomic add is emulated with a loop and an atomic compare-and-swap instruction (\texttt{ATOMS.CAS}), which are far more expensive than a single (native) global atomic add instruction (\texttt{RED.E.ADD.F32.FTZ.RN}). More details about the SASS instructions can be found in the instruction set reference~\cite{nvidia2021instruction}.}
    \label{fig:atomic_disassembly}
\end{figure}
We begin our discussion of these principles with the intrinsic functions. Some of the intrinsic functions are not directly mapped to a single hardware instruction. Instead, they are emulated by dozens of instructions. We name these functions as \emph{non-native} functions for brevity. For example, prior works~\cite{wang2020massively,hu2019taichi,gao2018gpu} rely heavily on atomic operations applied to floating-point shared memory variables. These operations are non-native and will be translated into complex instructions, e.g., branching and looping instructions (\figref{atomic_disassembly}), which can be more expensive. We demonstrate in~\secref{anna} that enforcing this principle by carefully redesigning the pipeline could deliver a substantial performance boost.

\begin{figure*}[htb]
\centering
    \includegraphics[width=1.0\linewidth]{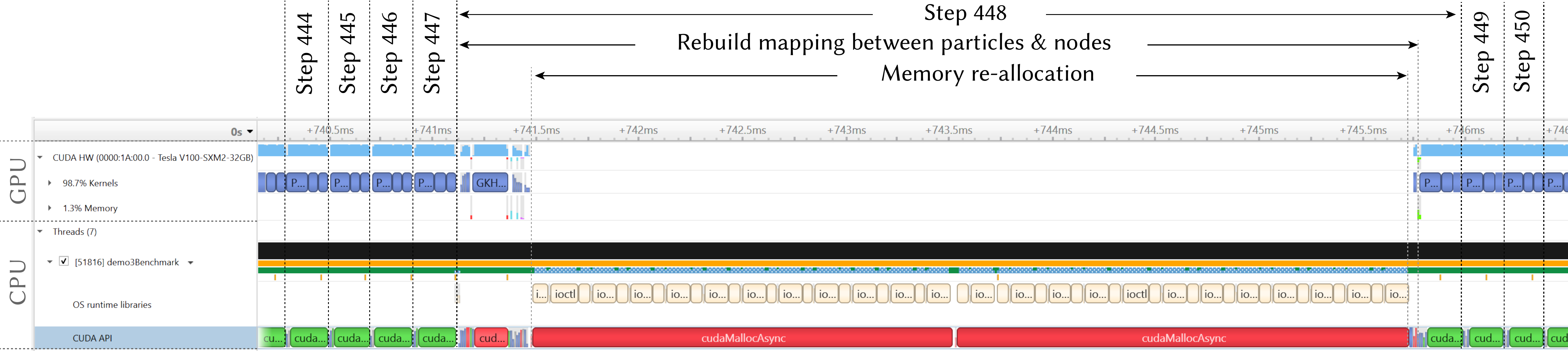}
    \caption{A screenshot of Nsight System analysis of an MPM simulation with 55K particles. Memory reallocations can take up a large proportion of a time step (even much larger than the actual computations), which is more likely to occur when the particles get stretched and cause the particle-node mapping to be rebuilt.}
    \label{fig:nsys_allocation}
\end{figure*}
The second principle is straightforward, and some prior works (e.g., Tarjan et al.~\shortcite{tarjan2009art}) have also emphasized this. More specifically, in scenarios with a small number of particles, without an optimized memory reallocation strategy, the memory reallocation (and the implicit synchronizations) would consume a longer time (\figref{nsys_allocation}) than the computational kernels themselves. 
Besides, it would also make the frame rate unstable and thus impact the user experience in interactive applications. However, it is also impractical to predict the necessary memory accurately or reserve all available memory for simulation. To minimize memory reallocation, we implement a GPU container similar to standard C++'s \texttt{vector}~\cite{stroustrup2013cpp} that would automatically allocate memory with $4\times$ of the requested size when it is half-filled. 

To maximize the performance, we minimize the number of CUDA kernels executed within a single time step, where two different ways can be applied. 
The first way is merging smaller kernels into a larger one. This idea is not new and has been discussed in Wang et al.~\shortcite{wang2020massively}. 
The second way is less evident than the first. We can divide the kernels per time step into two groups: the essential and non-essential groups. While the former is inevitable, kernels in the latter group are unnecessary to be conducted every step. In the context of MPM, we show in~\secref{rbmdetail} that the kernels rebuilding the mapping between particles and grid nodes are non-essential and can be invoked with a much lower frequency to achieve significant computational savings.

\begin{figure}[htb]
\centering
    \includegraphics[width=1.0\linewidth]{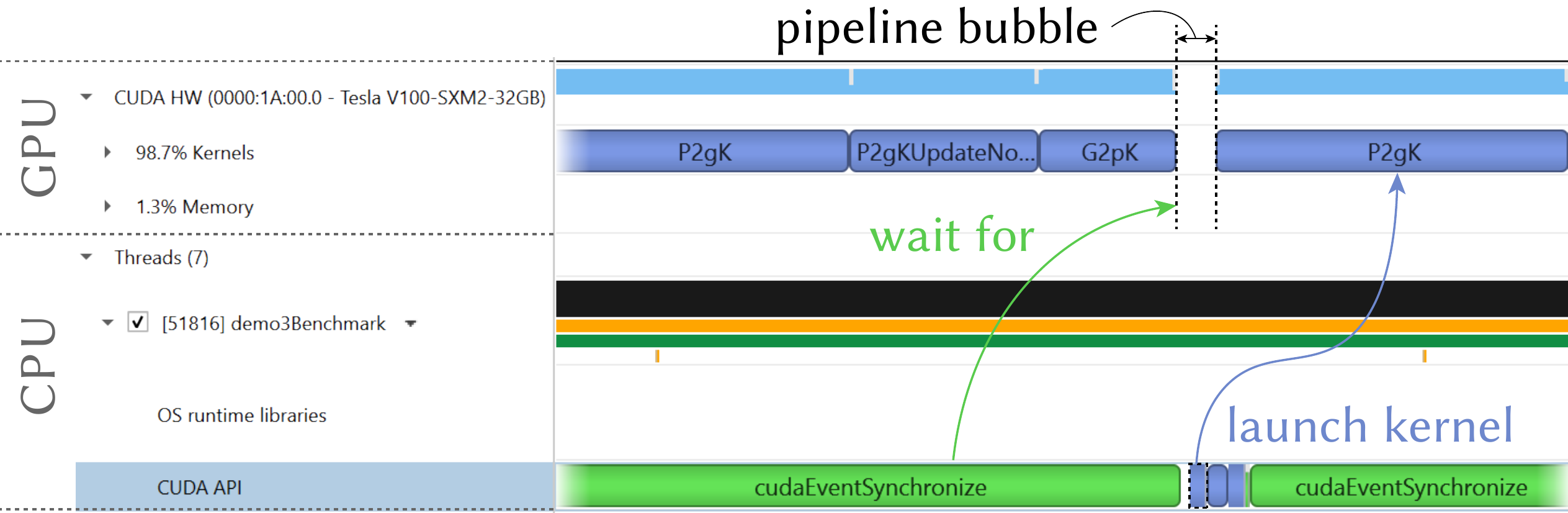}
    \caption{A screenshot of Nsight System analysis of an MPM simulation with 55K particles. A GPU-CPU synchronization performed through \emph{cudaEventSynchronize} waits for a G2P transfer to complete on the GPU, delaying the later kernel invocations on the CPU. As a result, a gap between the invocation and execution (or a pipeline bubble) appears.}
    \label{fig:nsys_gap}
\end{figure}
For most kernels, GPU and CPU work asynchronously. Issuing a kernel on the CPU usually takes much less time than executing it on the GPU, leading to a temporal difference between the invocation and execution. However, there are cases where a CPU-GPU synchronization is necessary. When there is a CPU-GPU synchronization, the time gap accumulated from the previous asynchronous executions would suddenly appear, causing the GPU to stall: the gap makes the GPU idling before the subsequent kernel execution (also known as a \emph{pipeline bubble},~\figref{nsys_gap}). 
We put as much work as possible on GPU. Besides, the necessary data transfer from GPU to CPU mostly happens when building data structures. For example, when building the particle-node mapping, the CPU code needs to know the number of particle blocks to determine how many CUDA threads need to be launched for the following kernels or if the memory needs to be reallocated. To reduce CPU-GPU synchronizations, we eliminate the rebuilding of such data structures as many as possible.

On a streaming multiprocessor (SM), the register file and shared memory size are limited. The more a CUDA block (containing several warps of threads) uses shared memory or the more each thread uses registers, the less number of CUDA blocks will be simultaneously active on the SM. As a result, the SMs will be less occupied, leading to worse efficiency. Therefore, we fine-tune the number of CUDA blocks used for each kernel and the balance between registers and shared memory usage. It is a common practice to cache data with shared memory. However, it is less well-known that spilling some registers into shared memory in a complex kernel (that may use too many registers by default) also helps increase occupancy (and thus efficiency). In the context of MPM, we spill the B-spline weights (nine \texttt{float}s for each thread) by storing them into the shared memory (instead of registers), although none of them is shared between threads. This choice increases performance, particularly when the number of threads is large (\secref{relieve_registers}).

\subsection{Multiple GPUs\label{sec:pcpmulti}}
Our principles proposed for a single GPU also apply to the case when simulated on multiple GPUs. For example, frequent memory reallocations cause even more severe overhead in a multi-GPU environment (\secref{reduce_alloc_multi}). Nevertheless, there are additional considerations for performance optimization in a multi-GPU environment.

Unbalanced work distribution among GPUs would detriment the performance. GPUs with less workload may stay idle while other GPUs keep busy, wasting the overall computing power.
In the ideal case, the workload would be dynamically divided into even slices and redistributed to GPUs every few steps. For example, in a smoke simulation, the workload covers several spatial blocks that change every step as the smoke rises, so the slices must be recalculated every few steps.
For MPM, as the computational workload is closely related to the number of particles, we divide the initial set into pieces with the equivalent number of particles based on their initial position. We then assign each piece of particles to one GPU and keep this partition unchanged throughout the simulation. We demonstrate that this simple partitioning scheme provides a balanced workload for scenes used in this paper.
However, dynamic particle partitioning should be introduced for extreme cases where particles from different GPUs can get evenly mixed, e.g., by rotating a paddle in the middle of a container. We leave it as future work.

We propose a few principles that maximize the performance of real-time applications on multiple GPUs. 
The first and most important principle is that one should always try to make each GPU process independently and only transfer information between different GPUs when inevitable.  
We further categorize all these principles into what and how to transfer among multi-GPUs for a more straightforward interpretation. A detailed ablation study on the impact of these principles is presented in~\secref{ablation_multi}.
\begin{itemize}
    \item Minimizing the number of transfers and synchronizations between GPUs;
    \item What to transfer?
        \begin{itemize}
            \item Minimizing the amount of data transferred between the GPUs and the subsequent computations;
        \end{itemize}
    \item How to transfer?
        \begin{itemize}
            \item Using in-kernel peer-to-peer (P2P) read/write operations for inter-GPUs communications;
            \item Overlap P2P data transfer and computation through warp-interleaved execution when using NVLinks
        \end{itemize}
\end{itemize}

The inter-GPU communications always appear with inter-GPU synchronizations. Communications are accomplished via either reading or writing between GPUs. When a GPU writes to some other GPUs, a post-write synchronization would be necessary to ensure the receivers get the information from the sender. Similarly, when a GPU reads from some other GPUs, a pre-read synchronization would help it know that all the data to be collected are in place. Concurrent reading and writing from/to the same address among GPUs should be avoided.

\begin{figure}[htb]
\centering
    \includegraphics[width=1.0\linewidth]{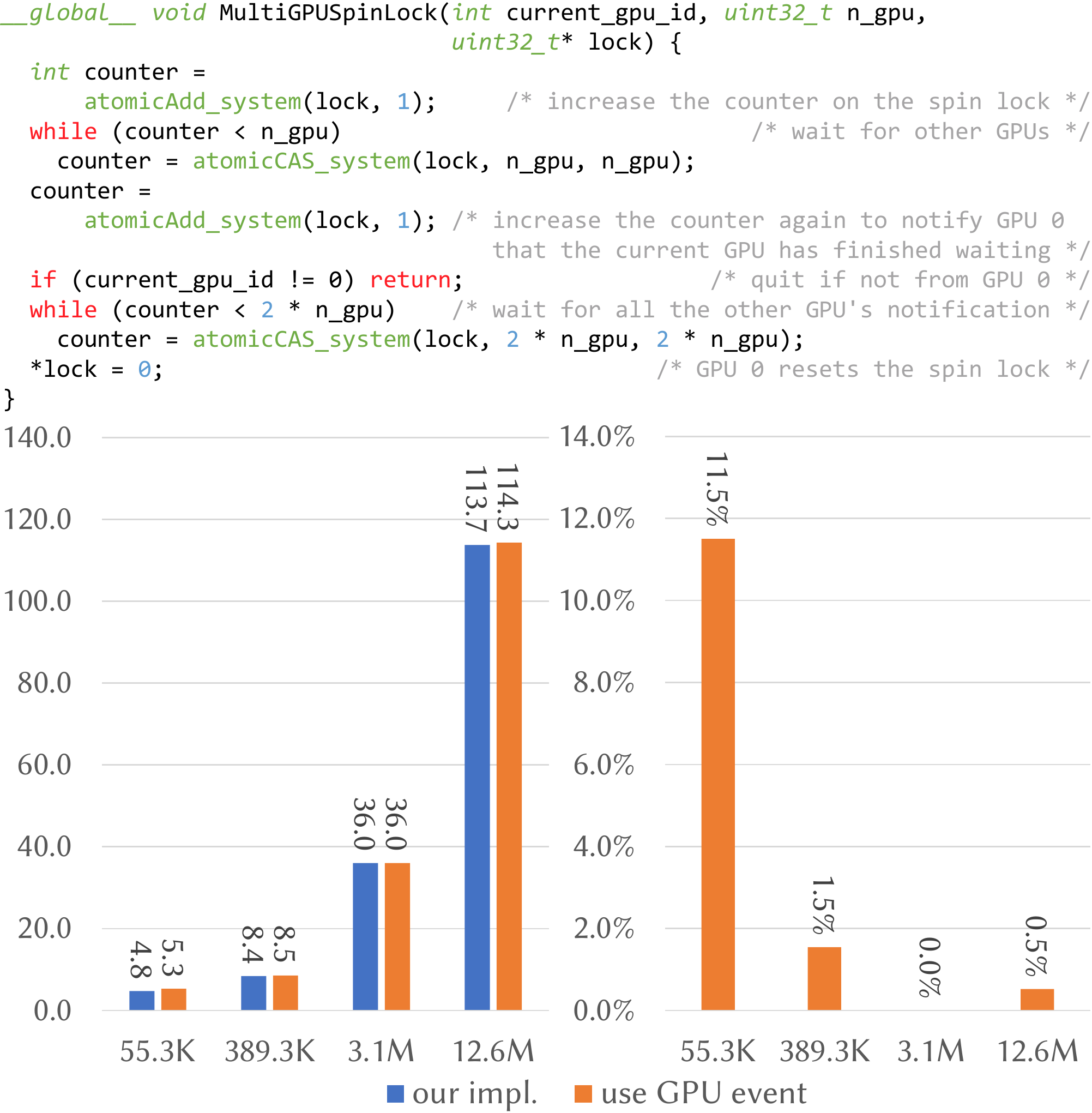}
    \caption{\textbf{Top:} A GPU spin lock kernel for the synchronization between different GPUs. The kernel is executed with a single thread on each GPU, the GPU id, number of GPUs, and a pointer to integer-sized device memory as the parameters. \textbf{Bottom:} Timing (in milliseconds) comparison between our implementation and the one with cudaStreamWaitEvent, as well as the percentage of additional costs incurred by the latter.}
    \label{fig:multi_gpu_mutex}
\end{figure}

The inter-GPU synchronizations usually involve locks between multiple streams and contexts which create pipeline bubbles on the GPUs, i.e., causing GPUs to stall.
Thus, in general, the number of synchronizations between GPUs should be minimized.
When synchronization is unavoidable, a common choice is directly using the \emph{cudaStreamWaitEvent} API.
However, as this API relies on sequential CPU mutexes to avoid deadlock, the cost would quadratically increase w.r.t. the number of GPUs.
Furthermore, it would be unacceptably expensive for scenes less than $100$k particles since the computations would not cover the cost.
We instead implement inter-GPU synchronization by creating a barrier across different GPUs via system atomics over NVLinks (or effectively, a \emph{spin lock}), and it does not need to have the CPUs involved. As shown in \figref{multi_gpu_mutex}, while the two schemes provide similar performance with a large number of particles, ours delivers much better efficiency with around $50$k particles.

The second principle emphasizes that the amount of data transferred between GPUs should be minimized. Each thread is in charge of a particle in the particle-grid transfer (P2G) of MPM. When threads from different GPUs write to the same set of nodes, the values of these nodes held by each GPU need to be combined to get the final values. 

There are two ways to combine the values from different GPUs. One way is to duplicate the data on particles and transfer them among GPUs before P2G. Then each GPU can process both the particles from itself and the transferred as usual without interfering with the other GPUs.
Another way is to apply a simple reduction to the grid nodes shared by multiple GPUs right after the P2G so that each GPU has the correct nodal values to move forward.
We adopt the second option as it transfers fewer data (i.e., the number of active nodes is usually much smaller than the number of particles, and nodes possess less data than particles) and saves from additional expensive P2G of extra copies of particles. In addition, operations that identify the particles and nodes shared among GPUs need to be performed when these particles' (or nodes') spatial distribution develops. Compared with the first option, the tagging scheme used in the second option is much more lightweight and brings much less overhead.

The last set of principles consists of two aspects about how exactly the inter-GPU transfers should be achieved. 
For transferring data among GPUs, specifically for scenes with a small number of particles, in-kernel peer-to-peer (P2P) accesses should be preferred than dedicated copy engines such as a bunch of \emph{cudaMemcpy}-s or the NCCL library~\cite{jeaugey2017nccl}, as the latter brings in an extra invocation cost. Furthermore, SMs on modern GPUs can efficiently switch between executing warps waiting for a P2P transfer and warps conducting computations. Therefore, we fuse the P2P copying with some computations into a single kernel and rely on GPU's interleaved warp execution to overlap data transfer and computations. 

\paragraph{Remark:} The last one seems to be controversial as it is almost a standard way in high-performance computing (HPC) to hide device-to-device communication costs via manually splitting data transfer as an independent stream and overlapping it with computations~\cite{wang2020massively}. However, multiple streams would work best if the communications consume roughly the same amount of time as the computations or within specific copy engines that have logic control units running in parallel with the computational units (e.g., MPI for a CPU-GPU heterogeneous framework~\cite{dominguez2013new}).
For (nearly) real-time simulations, the situation is different. The total amount of data on each GPU is much smaller than the amount that needs to be considered in HPC applications in the first place. Furthermore, only a tiny portion of it needs to be transferred between GPUs, making the communication cost only a tiny fraction of computational cost. The P2P data transfer, although very efficient, also occupies SMs. Thus, manually splitting a kernel into two kernels, i.e., a P2P kernel and a computation kernel, would not be beneficial.

\section{Material Point Method on a Single GPU}
\begin{figure}[htb]
\centering
    \includegraphics[width=1.0\linewidth]{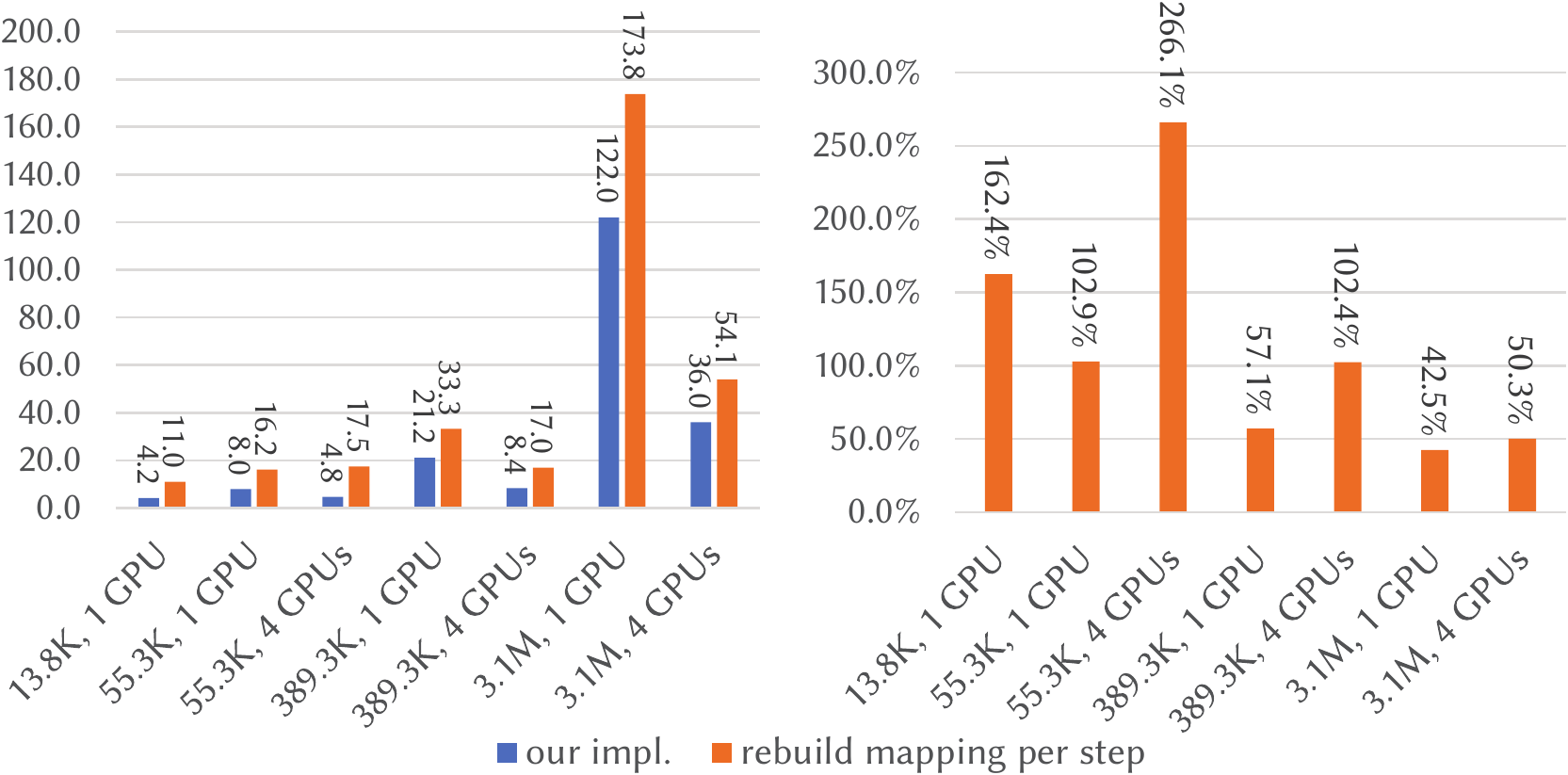}
    \caption{\textbf{Left:} Timing (in milliseconds) comparison between our implementation and the one where mapping between particle and grid are rebuilt every step; \textbf{Right:} The percentage of additional cost incurred by rebuild-mapping per step.}
    \label{fig:single_gpu_rbm}
\end{figure}
Some prior works~\cite{wang2020massively,hu2019taichi,gao2018gpu} also focus on optimizing the material point method on GPUs.
All of them share a similar procedure that contains the following phases for each time step:
\begin{enumerate}
    \item \textbf{Rebuild-mapping}: building up the mapping between particles and background grid blocks;
    \item \textbf{Sort-particle}: sorting particles to the appropriate level of granularity: the level of cells~\cite{gao2018gpu} or the level of blocks~\cite{wang2020massively,hu2019taichi});
    \item \textbf{Transfer-particle-to-grid}: tranferring particle momenta and masses to grid nodes and computing nodal forces;
    \item \textbf{Update-grid}: updating nodal velocities and resolving collisions on grid nodes;
    \item \textbf{Transfer-grid-to-particle}: transferring new velocities back to particles from grid nodes and advecting particles.
\end{enumerate}
The rebuild-mapping phase consists of several substeps. First, we compute particle codes from particle positions, and we adopt a simplified version of (GPU) SPGrid \cite{setaluri2014spgrid,gao2018gpu} that interleaves the three-dimensional 32-bit indices of particles into a 64-bit offset. The higher bits encode the block information (\emph{block code}) while the lower bits encode the cell information (\emph{cell code}).
With the particle codes, we construct a hash table to record all blocks in which particles reside (named as a geometric block~\cite{gao2018gpu}, or \emph{gblock}). Next, we update the hash table by recording all the $3\times 3\times 3$ neighboring blocks of each geometric block. These blocks, including the gblocks and their neighboring blocks, cover all the nodes the particles may talk to through B-spline quadratic weighting functions. We name this group of blocks as physical blocks (\emph{pblock}), of which the gblocks are a subset.
These pblocks are sparsely populated in space while they are consecutively stored and indexed in memory. In this work, the block code would have a sense of space while the block index is used to record the block's memory location; and we update the mapping between them during the rebuild-mapping phase.

As the shared memory has a limited size on streaming multiprocessors, prior works~\cite{gao2018gpu,wang2020massively,hu2019taichi} require that the particles handled by a CUDA block should come from the same particle block so that the amount of shared memory allocated to each CUDA block can be reasonably small.
To achieve this, Gao et al.~\shortcite{gao2018gpu} partition the particles from the same particle block into a few CUDA blocks and name such CUDA blocks as ``virtual blocks'' (or \emph{vblock}). 
In contrast, as our pipeline gets rid of shared memory (refer to~\secref{anna} for details), we divide particles by the warps. We only require the particles in the same warp are originated from the same particle block in the rebuild-mapping phase, and each CUDA block can handle warps from different particle blocks. 

There are several choices of algorithms for sorting particles. Radix-sort using particle codes as keys is the simplest. However, as discussed by Gao et al.~\shortcite{gao2018gpu}, a histogram-sort performs more efficiently, where the keys are computed through concatenating the block index and the cell code. Therefore, we adopt the latter for sorting particles. Unlike prior works~\cite{wang2020massively,hu2019taichi} that only sort the particles to the granularity of blocks, in this work, we sort the particles to the granularity of cells. There are two benefits of this choice. First, processing particles that are sorted w.r.t. cells during the P2G and G2P transfer is more efficient since we can read from or write to consecutive memory (see~\secref{result_sp} for validation). Second, it enables using the warp-level reduction before atomically writing data to nodes~\cite{gao2018gpu} for expedited particle-to-grid (P2G) transfers, which leads to much less atomic operations compared with works using block-level sorting.

The P2G kernel also computes each particle's plasticity, stress, and body force before conducting the actual P2G transfer. After P2G, the computations on the grid nodes take the collisions with boundaries into consideration to update the nodal velocities. Then the final grid-to-particle (G2P) phase transfers the velocities back to particles.

\subsection{Merge Kernels}
\begin{figure*}[htb]
\centering
    \includegraphics[width=1.0\linewidth]{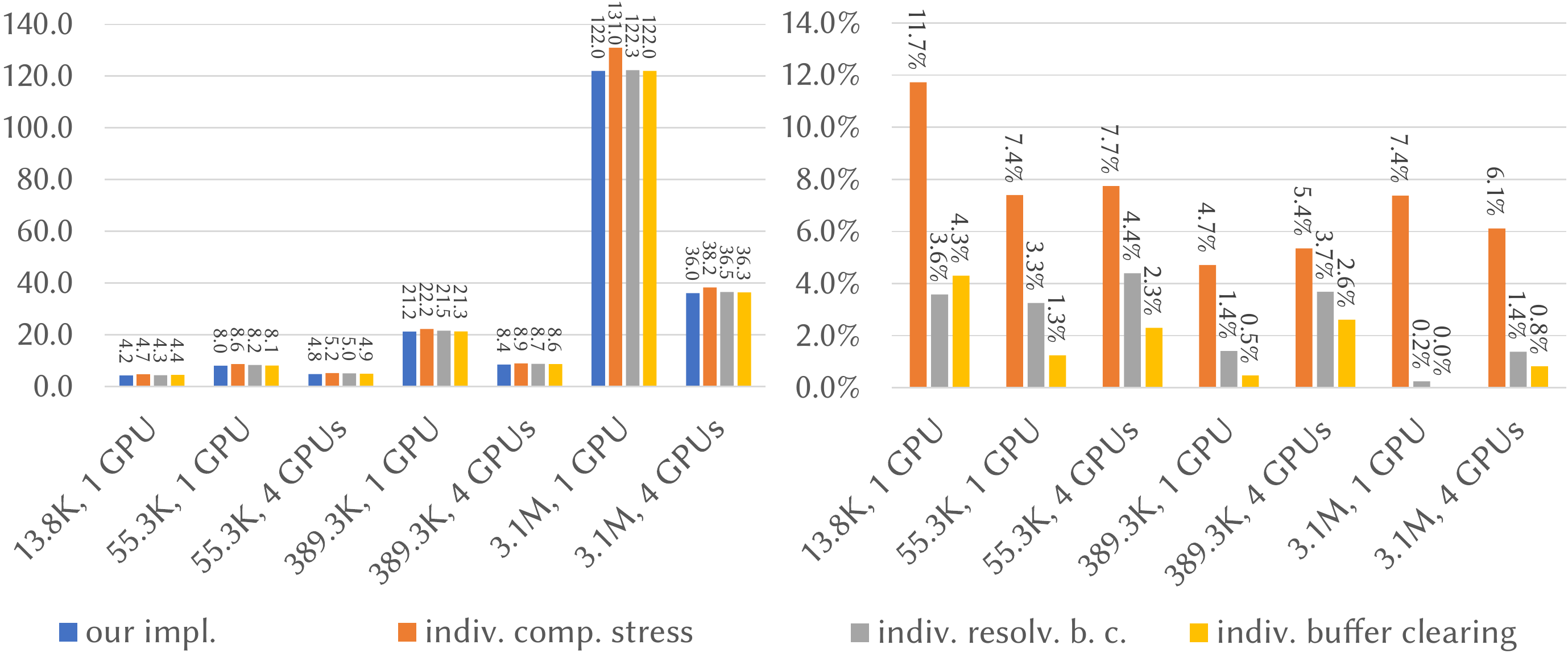}
    \caption{\textbf{Left:} Timing (in milliseconds) comparison between our implementation, the one invoking an individual kernel to compute stress before P2G, the one invoking an individual kernel to resolve boundary conditions after P2G, and the one invoking an individual kernel to clear the nodal buffer before P2G; \textbf{Right:} The percentage of additional costs incurred by different designs other than ours.}
    \label{fig:single_gpu_indiv}
\end{figure*}

The key benefit of merging kernels is that specific global memory accesses can be avoided for the variables that are not part of the system state. For example, the stresses computed on particles are temporary states used to compute the nodal forces and do not need to be stored in the global memory for the next step, and therefore, we can compute the stresses with P2G in the same kernel.

This idea of merging kernel is not new and first employed by Hu et al.~\shortcite{hu2019taichi} that they postpone the update of deformation gradient at the end of G2P to the beginning of P2G to avoid the redundant write operations. Wang et al.~\shortcite{wang2020massively} move one step further to merge quite a lot of kernels into a single large kernel - the G2P2G kernel.

Nevertheless, a complicated kernel may also introduce additional costs or limitations in certain situations. For example, it would be challenging to simulate Lagrangian MPM models (e.g., hair or cloth~\cite{jiang2017anisotropic,guo2018material,fei2021revisiting}) with a G2P2G kernel, as some of these models require sophisticated modifications to the particles' velocities before P2G. Also, operations such as the addition and deletion of particles are not easily supported in G2P2G as they usually happen between G2P and P2G, forbidding particle sources or dynamic particle boundaries~\cite{stomakhin2017fluxed,huang2021ships}. In the result section, we have also conducted one experiment to illustrate that combining G2P2G with our current pipeline would only deliver better performance when the number of particles is relatively small. We observed that the compiled SASS code of G2P2G is extended and contains some long-distance jumps. When there are many particles so that there are many warps to be executed, the execution between warps can vary, requiring a large instruction cache. Because the latter is limited, a dramatic increase of instruction cache miss is observed, which can be justified by a noticeable rise of \emph{stall no instruction} in Nsight Compute analysis.

To avoid these drawbacks of G2P2G, we separate the two transfer phases unless there are no particle additions or deletions and the number of particles is small (e.g., less than 100K in this work). However, we merge all computations on particles, i.e., singular value decomposition, plasticity projection, stress computation, into a single P2G phase. 

Another benefit that is less discussed is that by merging small kernels into an appropriate size, the stream multiprocessors (SMs) can overlap memory accesses with computations for better performance through warp scheduling. The latter is efficient and costs only one cycle to switch between two warps~\cite{jia2019dissecting}. 
For example, the P2G kernel overlaps the writing operations at its end with the computations at its beginning.

Besides, when the simulation resolution is small, there are only a limited number of thread blocks. The SM groups thread blocks into a specific size (known as a \emph{wave}) and execute the thread blocks wave by wave. Without merging kernels, the thread blocks would not have enough computations; the limited number of threads in the last wave makes SM idling, wasting a large portion of computational power. Such inefficiency is also known as the \emph{tail effect}~\cite{micikevicius2012gpu}.
With merging, these trivial computations can be executed and overlapped with other heavier computational or memory operations, eliminating the tail effect and drastically improving the overall performance.~\footnote{When the simulation resolution is large, the tail effect will fade out as the number of waves becomes large. Hence, the discussion here only applies to the case where simulation resolution is small.} 

\subsection{Minimize nonessential computations}
As mentioned in the previous section, the computations per time step can be divided into two groups. Specifically, for the five phases of the material point method, we regard the first two as the nonessential ones and the other three as the essential ones. While the essential phases would always stay in place, the nonessential phases can sometimes be removed to save the cost of performing them. One may naively remove them to ignore rebuild-mapping and use a dense background grid or remove the particle sorting not to concern the performance of the memory operations. Such naive ways, however, would always bring some other side effects that impact the overall performance.

Instead, in this work, we develop a new method that vastly reduces the frequency of these two nonessential phases to achieve substantial computational savings without bringing other side effects. In other words, instead of running them every time step, we can now conduct them once every few time steps with appropriate modifications.

\subsubsection{Rebuild-Mapping\label{sec:rbmdetail}}
\begin{figure}[htb]
\centering
    \includegraphics[width=1.0\linewidth]{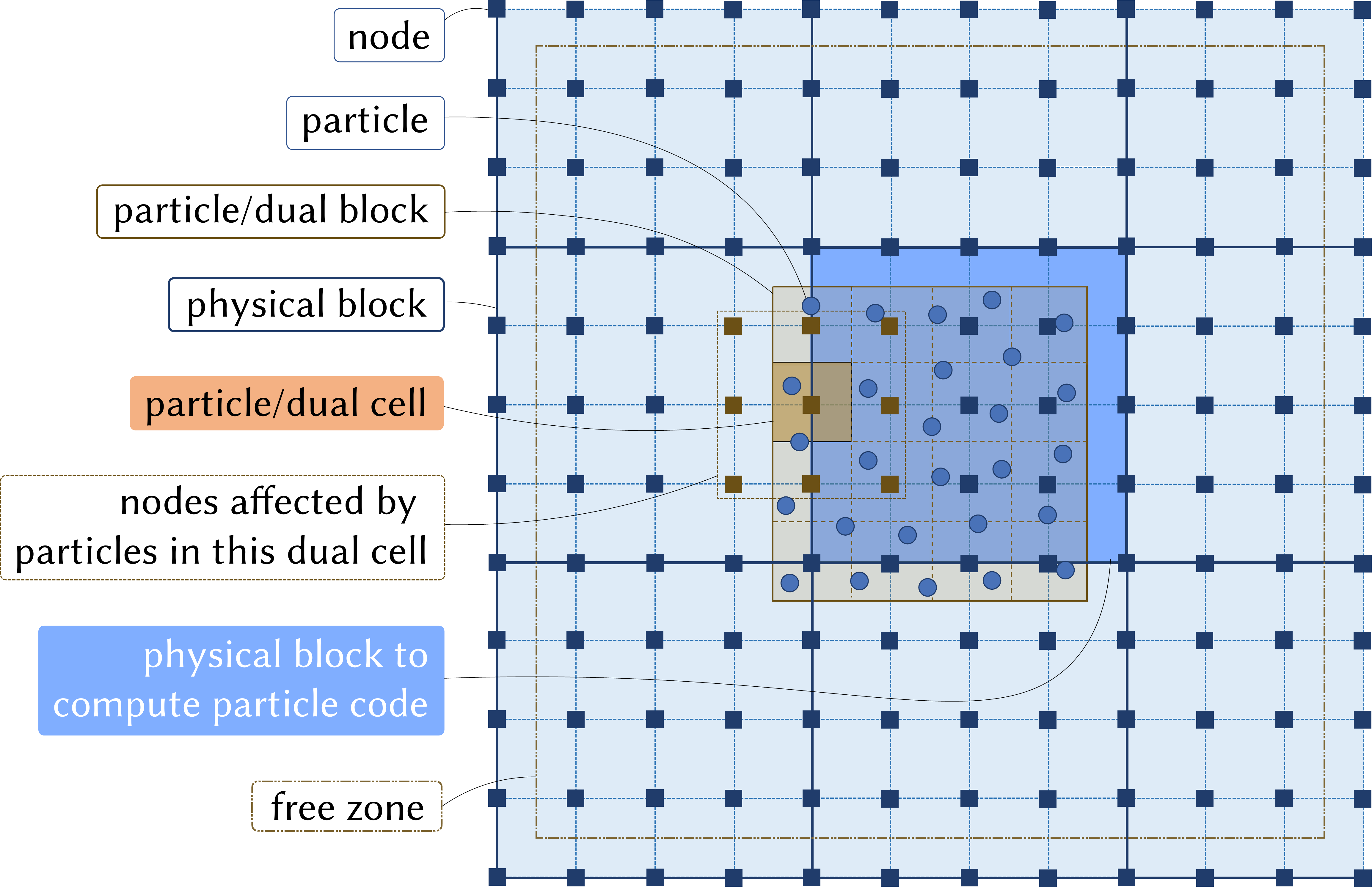}
    \caption{\textbf{Mapping between particles, nodes, and physical blocks.} For the particles in a particle block, we compute their codes through the physical block that has $0.5\text{dx}$ shifting from the particle block during rebuild-mapping. Without triggering another rebuild-mapping, these particles can move in the \emph{free zone} and affect (potentially) $12\times 12$ (or $12\times 12\times 12$ in 3D) nodes in $3\times 3$ (or $3\times3\times3$ in 3D) physical blocks.}
    \label{fig:rebuild_mapping}
\end{figure}
We start with the scheme proposed in Gao et al.~\shortcite{gao2018gpu} to rebuild mapping between the particles and nodes: particles are initially divided/sorted into non-overlapping particle blocks. As shown in Fig. 4 of Gao et al.~\shortcite{gao2018gpu}, a particle block is shifted by half-dx w.r.t grid block, such that the particles assigned to this particle block will touch and will only touch the nodes of its $2\times2\times2$ neighboring blocks. A particle's assigned block can then be found by right-shifting its code only to keep the higher bits, and the neighboring blocks can also be retrieved either via the hash table or via an explicitly stored vector. In the next step, as particles advect, some may end up with positions outside their original particle block and need to talk to the nodes that are not tracked. Thus, rebuilding the mapping between particles and grid blocks is inevitable.

We observe that the aim of rebuild-mapping is not to find the grid block in which a particle resides but rather to provide the memory addresses of the neighboring 27 nodes with which the particle will interact through a quadratic kernel. If the moving range of a set of particles is known and fixed in a few steps, one may find all nodes that possibly interact with these particles and build the mapping between them \emph{once for all these steps}.

In this work, we allow the particles to move within a $(10\text{dx})^3$ range (named as a \emph{free zone}, see~\figref{rebuild_mapping} for an illustration) without triggering rebuild-mapping. Instead of $2\times 2\times 2$ neighboring blocks, we divide particles into blocks with a negative-half-dx shifting and keeping track of its $3\times 3\times 3$ neighboring blocks. As long as the particles move in the \emph{free zone}, they are free from rebuild-mapping. In other words, all grid nodes with which the particles communicate are known, and their addresses can be computed from the blocks' codes. 

Our scheme can substantially reduce the frequency of rebuild-mapping. The CFL condition usually prevents particles from traveling a distance longer than the cell size dx within a time step, and thus using a free zone with $10\text{dx}$-width may potentially reduce the frequency of rebuild-mapping from every step to per four steps (assuming the extreme case, e.g., a particle on the edge of a dual block moves away from the block with one dx per step). In our test case where each frame contains 36 steps (\secref{results}), however, we observed the rebuild-mappings happen only every $10$--$30$ steps, and each frame only needs $1$--$4$ rebuild-mappings.

Wang et al.~\shortcite{wang2020massively} proposed a similar idea to allow transferring particle properties to grid nodes after advection within the G2P2G step as it is impossible to rebuild the mapping inside a kernel. However, they still conduct the rebuild mapping every time step.
They use $2\times2\times2$ blocks with one-and-half dx shifts, while our free zone is designed with a larger size to reduce further the frequency of rebuilding.
One can push to use more neighboring blocks than we do, and the extreme would end up with a dense background grid, where the rebuild mapping can be removed entirely.

Our scheme takes a longer time to hash more neighboring blocks. However, the hashing only happens during rebuild-mapping. Because we only rebuild mapping over a few steps, the cost of the entire rebuilding is amortized to these steps and becomes negligible. Additionally, to avoid handling more physical blocks, we mark the physical blocks that are not touched by any particles and skip them during the grid computations. In our experiments (\figref{single_gpu_rbm}), when the number of particles is small, i.e., $55.3$k, the rebuild-mapping itself is the bottleneck, and our free zone scheme alone brings $3.7\times$ acceleration. 

\subsubsection{Particle Sorting}
\begin{figure}[htb]
\centering
    \includegraphics[width=1.0\linewidth]{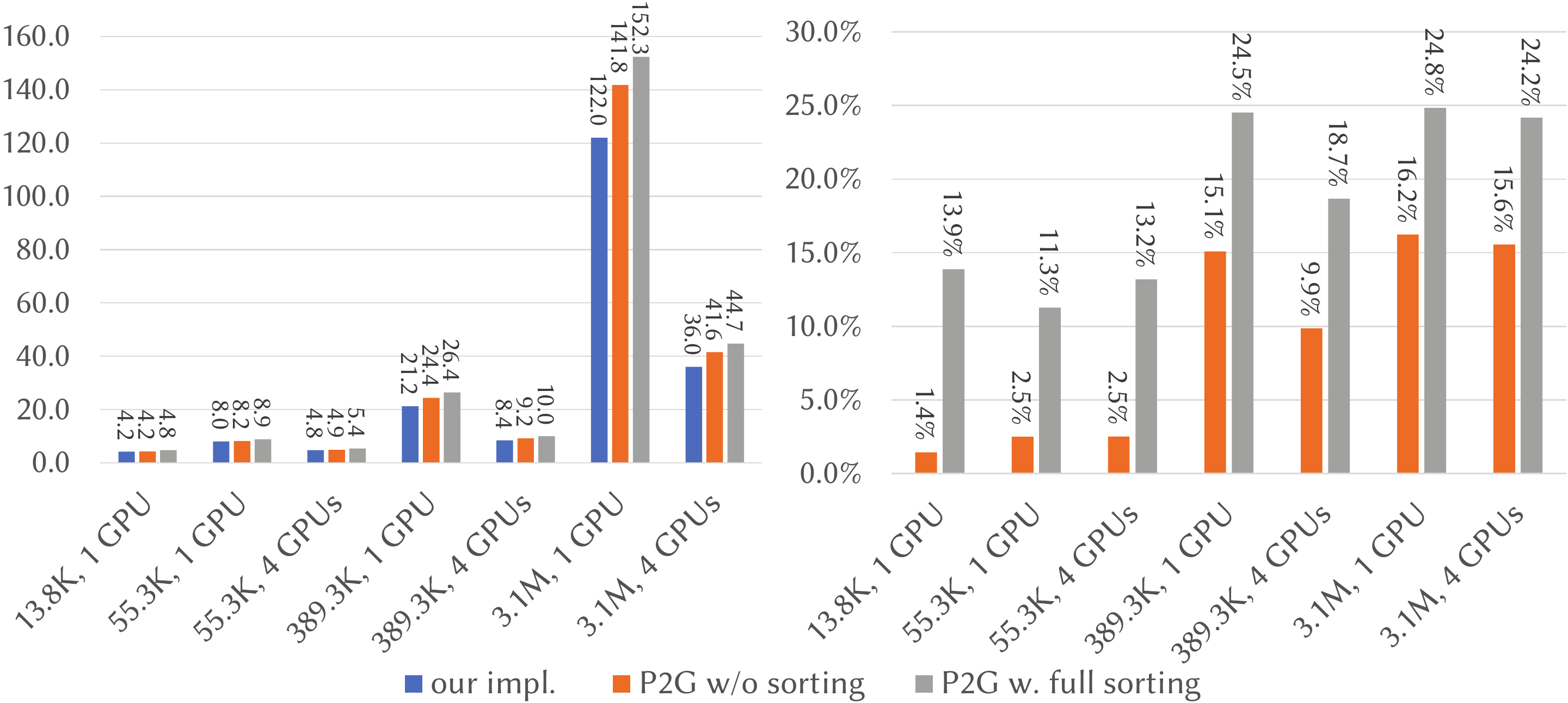}
    \caption{\textbf{Left:} Timing (in milliseconds) comparison between our implementation, the one without particle sorting, and the one re-sort all particles every step; \textbf{Right:} The percentage of additional costs incurred by not sorting and full re-sort.}
    \label{fig:single_gpu_sorting}
\end{figure}
In the particle-to-grid step, particles in the vicinity could simultaneously write to the same grid nodes, leading to writing conflicts. The simplest way to enforce correctness is to use CUDA atomic operations; however, the performance can be severely compromised without careful treatment. Two different types of solutions have been proposed to reduce the total number of atomics at any instant. 
Prior works~\cite{hu2019taichi,wang2020massively} either leave particles unsorted at the cell level or adjust their order with shuffling or binning strategies, such that particles within the same warp would have a smaller chance to conflict. 
On the contrary, Gao et al.~\shortcite{gao2018gpu} sort particles into cells and then apply a warp-level reduction to fuse multiple atomic operations into one.
We adopt the second way as it is more robust towards extreme scenarios. For example, when particles get tightly compressed into one cell, the first method would fail to alleviate conflicts.

\begin{figure*}[htb]
\centering
    \includegraphics[width=1.0\linewidth]{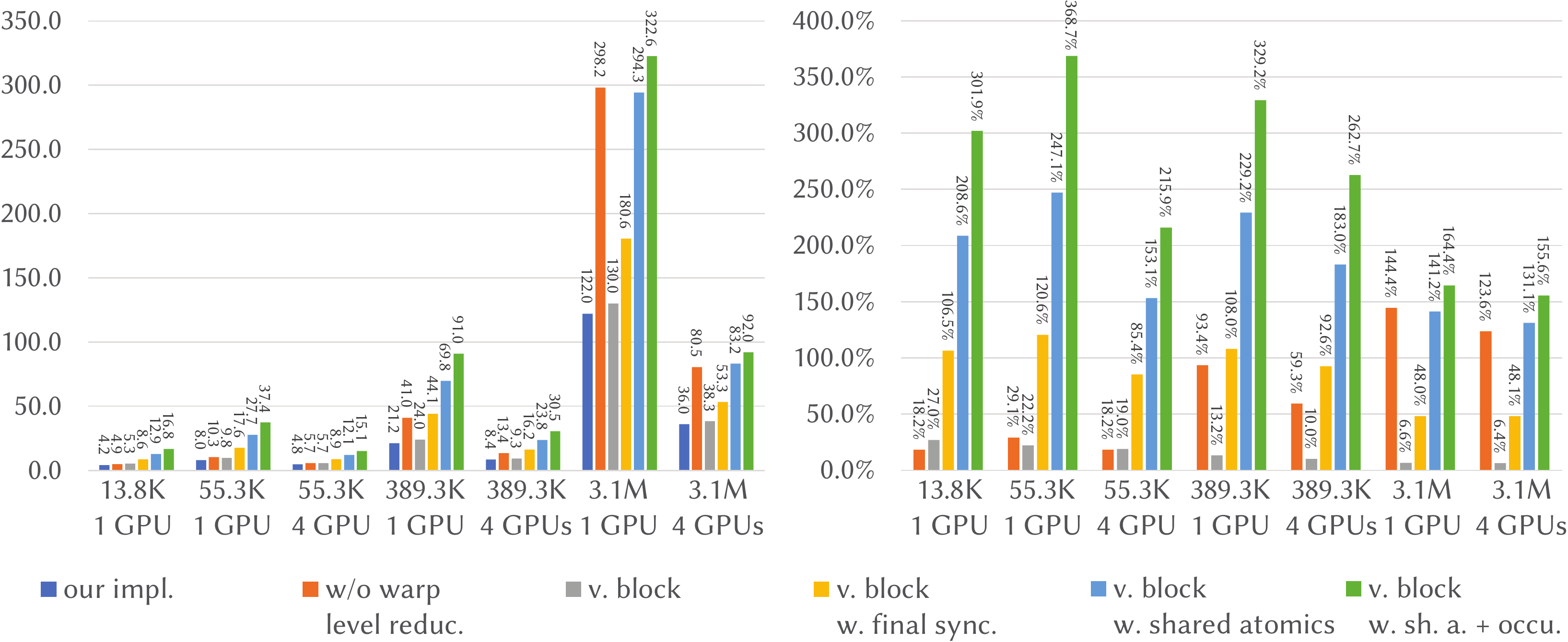}
    \caption{\textbf{Left:} Timing (in milliseconds) comparison between our implementation, the one without warp level reduction, the one using virtual block proposed in Gao et al.~\shortcite{gao2018gpu}, the one using virtual block and having an intra-block synchronization at the end of P2G, the one using virtual block with shared memory atomics, and the one using virtual block with shared memory atomics and higher occupancy; \textbf{Right:} The percentage of additional costs incurred by different designs other than ours.}
    \label{fig:single_gpu_vblock}
\end{figure*}
In a CUDA warp, 32 threads handle 32 particles in parallel. When particles are randomly ordered, the 32 threads would have a better chance to write to different grid nodes at one instant. However, as we sort the particles into cells, several subgroups usually exist among the 32 particles; threads inside each subgroup write to the same node. Reductions are independently conducted per subgroup, and then the first thread in each subgroup is responsible for writing the reduced result back to the grid node. In this way, each subgroup ends up with only one atomic operation. 

A complete particle sorting can be divided into two phases: sort the particles w.r.t. particle blocks and then sort w.r.t. cells.
We modify the first one to be compatible with the free zone, i.e., as long as the particles stay inside the free zone, there is no need to sort the particles into blocks.

Furthermore, as the reduction only helps to lessen the atomic operations within each warp, instead of sorting w.r.t. cells every time step, we can perform it only when rebuild-mapping happens. Between two rebuild-mappings, we conduct radix sort in each warp before the reduction in P2G transfer.
This arrangement splits sorting w.r.t cells further into two substeps: sorting in each warp and then sorting between warps, and the latter can be avoided in most time steps.
Notice that instead of using 32 bits for warp-level radix sort, we only need 10 bits for better efficiency as our free zone spans a domain with $(10 \text{dx})^3$.

Our new scheme may present a less optimal particle ordering, e.g., particles in the same cell can be distributed to several warps, resulting in several atomics instead of one. However, this performance loss can be compensated well when particle density is not extremely high in each cell. We avoid the complete particle sorting, which may possess quite a fraction of the whole pipeline, leading to $8\%$--$17\%$ acceleration across different resolutions.

\subsection{Avoid non-native intrinsics\label{sec:anna}}
Prior works~\cite{gao2018gpu,hu2019taichi,wang2020massively} adopt two types of atomic operations in the P2G step. Values such as weighted momenta transferred from particles to grid nodes are first added up in the shared memory by non-native atomics (\figref{atomic_disassembly}) to get intermediate results. They are then written back to global memory through native atomics to reach the final results.
However, non-native atomics are much slower than the native ones as they are not supported by the (latest generation of) hardware. Atomics of floating-point numbers performed on shared memory are emulated by a read-modify-write (RMW) operation using the \emph{atomicCAS} instruction.

Furthermore, the adoption of shared atomics also imposes constraints that may contradict our rebuild mapping scheme. The shared memory needs to be sufficiently large to store the data of all nodes with which the particles in a block may interact. However, as the streaming multiprocessor (SM) has a limited size of shared memory, kernels using a large shared memory will be executed with fewer active CUDA blocks. To make the required shared memory size acceptable, prior works~\cite{gao2018gpu,hu2019taichi,wang2020massively} limits the size of neighboring blocks as $2\times2\times2$ and require each CUDA block only to handle particles from a specific grid block. Instead, our rebuild mapping scheme uses $3\times 3\times 3$ neighboring blocks, which would require a shared memory size that significantly reduces the number of active CUDA blocks, leading to a dramatic performance loss.

Following the principle of avoiding non-native atomics, we directly write back to global memory via native atomics when computing transfers from one particle to one grid node. We achieve $3\times$-$5\times$ acceleration (\figref{single_gpu_vblock}).
Besides, $3\times3\times3$ blocks are now compatible with the free zone without fretting the memory size limitation.
The multiprocessor occupancy is also improved by allowing a CUDA block to handle warps from different particles blocks.
Furthermore, there is no need to synchronize threads before writing from scratchpad to global memory anymore. Warps that finish their work sooner would not need to stall but immediately switch to new tasks.
Similarly, we also get rid of the shared memory in the G2P step.

\begin{figure}[htb]
\centering
    \includegraphics[width=1.0\linewidth]{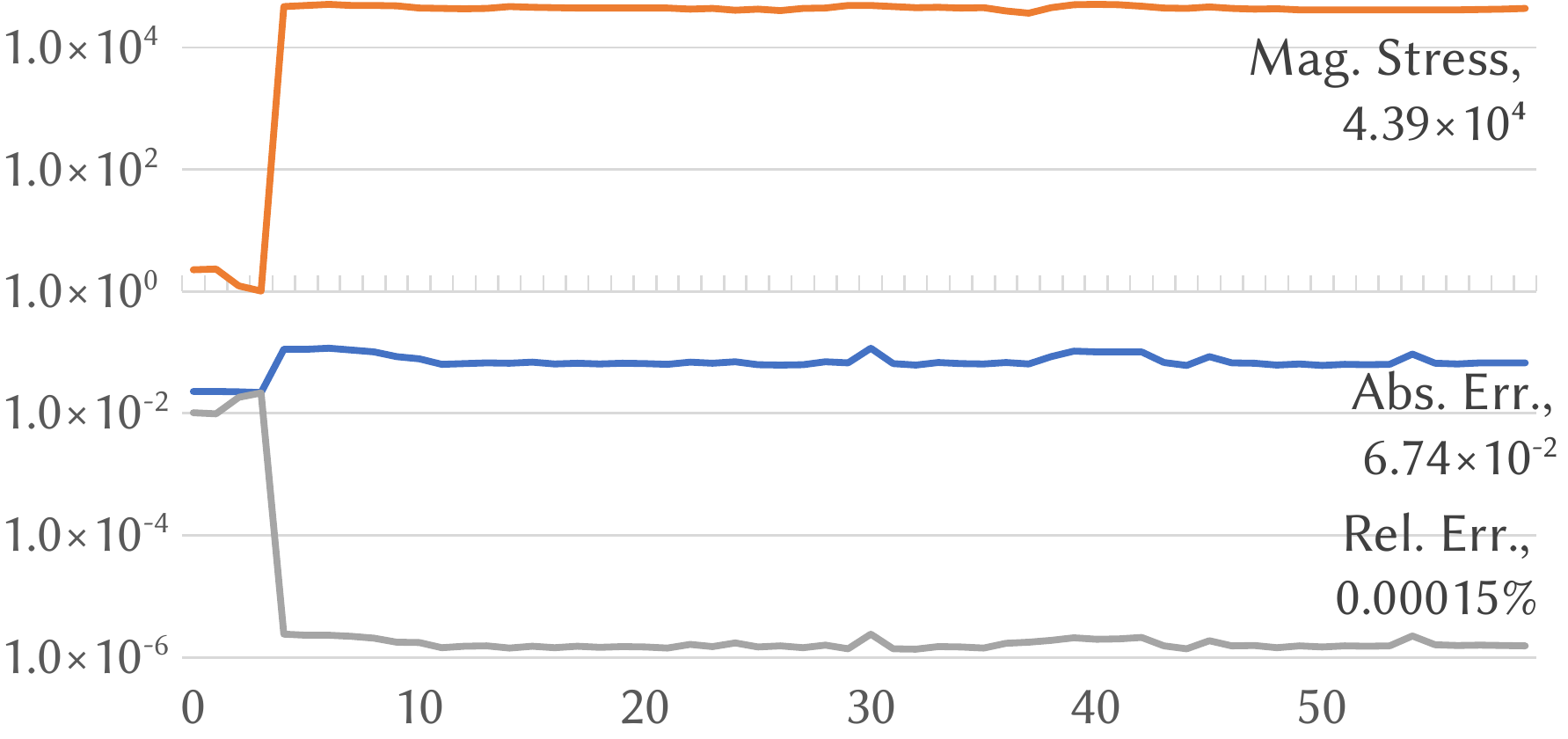}
    \caption{\textbf{Error of stress over different frames.} We use the binary without fast-math intrinsics as the baseline. \emph{Mag. Stress} is the infinite-norm of the tensor $\|\boldsymbol{\tau}_{\text{baseline}}\|_{\infty}$ in which $n$-particles' Kirchhoff stresses are stacked (i.e., $\boldsymbol{\tau}_{\text{baseline}}\in\mathbb{R}^{n\times 3\times 3}$) and computed with the baseline version; \emph{Abs. Err.} is the absolute error between the tensors computed with and without fast-math intrinsics, i.e., $\|\boldsymbol{\tau}_{\text{fast-math}}-\boldsymbol{\tau}_{\text{baseline}}\|_{\infty}$; and \emph{Rel. Err.} is the relative error between the two tensors, i.e., $\|\boldsymbol{\tau}_{\text{fast-math}}-\boldsymbol{\tau}_{\text{baseline}}\|_{\infty}/\|\boldsymbol{\tau}_{\text{baseline}}\|_{\infty}$. Numbers for the last frame are shown. The unit of the norm and absolute error is $\text{dyne}/\text{cm}^2$. The sudden jolt of magnitude and errors happens when the sand particles touch the floor and start to spread.}
    \label{fig:fast_math_error}
\end{figure}
Besides atomic functions, some other non-native intrinsics should also be avoided when possible. 
In CUDA, by default, the floating-point operations like $\frac{x}{y}$, $\text{sinf}(x)$ and $\text{logf}(x)$, are designed to match IEEE 754 precision standard \cite{IEEE754}.
To achieve this precision, the compiler will map these math functions to dozens of instructions. 
In cases where precision is not critical, the \texttt{-use\_fast\_math} compile flag can be used to utilize native intrinsic functions, which can save a lot of instructions and register costs (refer to~\secref{result_fast_math} for detailed comparisons). In~\figref{fast_math_error}, we compare the stress computed with and without fast-math intrinsics over 60 frames in the case with 3.1M falling sand particles (refer to~\figref{compare_wang}). The error brought by fast-math intrinsics is negligible (the absolute error is at most $0.12\text{dyne}/\text{cm}^2$), especially when the stress itself has a considerable magnitude. In practice, we did not observe any noticeable visual difference between the two simulations.

\begin{figure*}[htb]
\centering
    \includegraphics[width=1.0\linewidth]{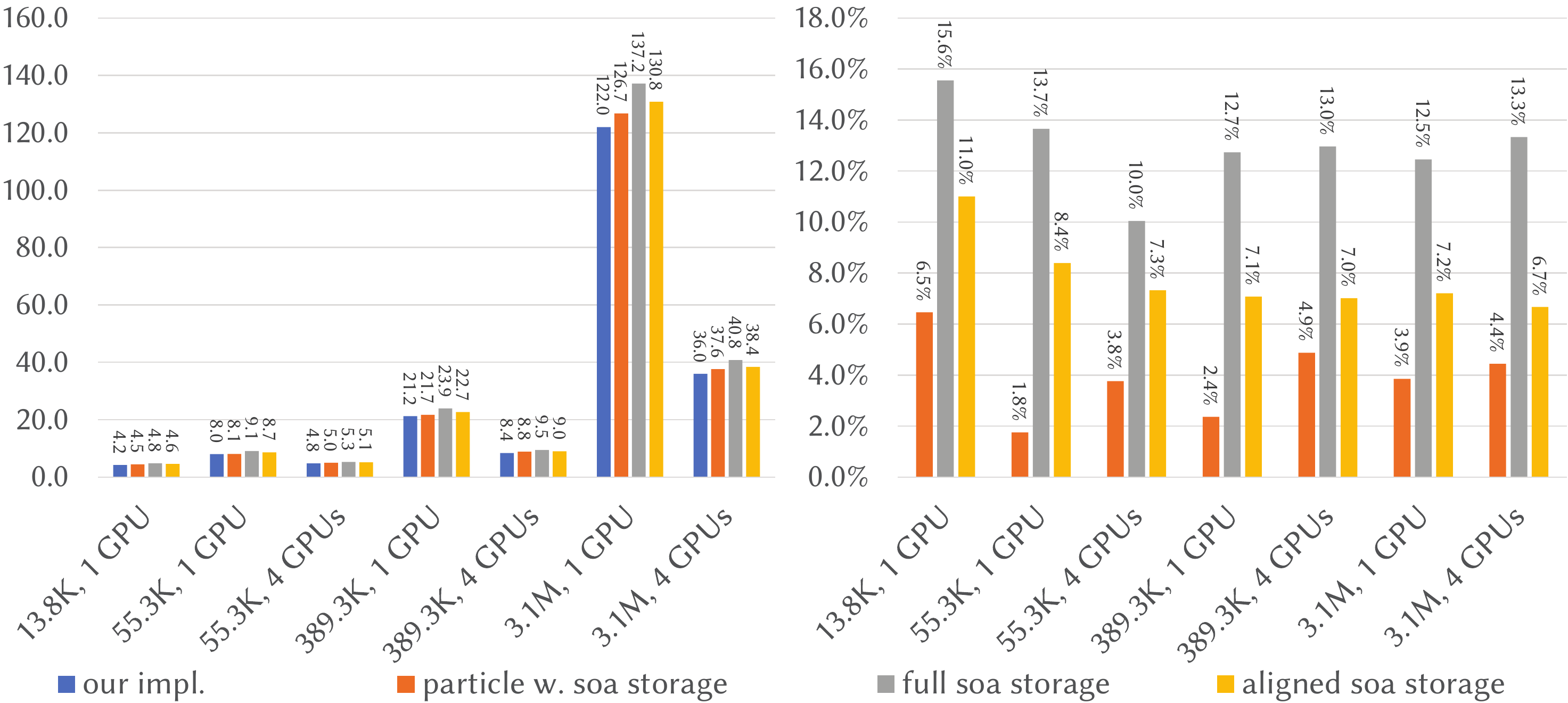}
    \caption{\textbf{Left:} Timing (in milliseconds) comparison between our implementation, the one uses SoA for particle storage, the one uses SoA for both particle and node storage, and the one uses SoA storage but aligned to 256 bytes; \textbf{Right:} The percentage of additional costs incurred by different designs other than ours.}
    \label{fig:single_gpu_storage}
\end{figure*}

\section{Material Point Method on Multiple GPUs}
\begin{figure}[htb]
\centering
    \includegraphics[width=1.0\linewidth]{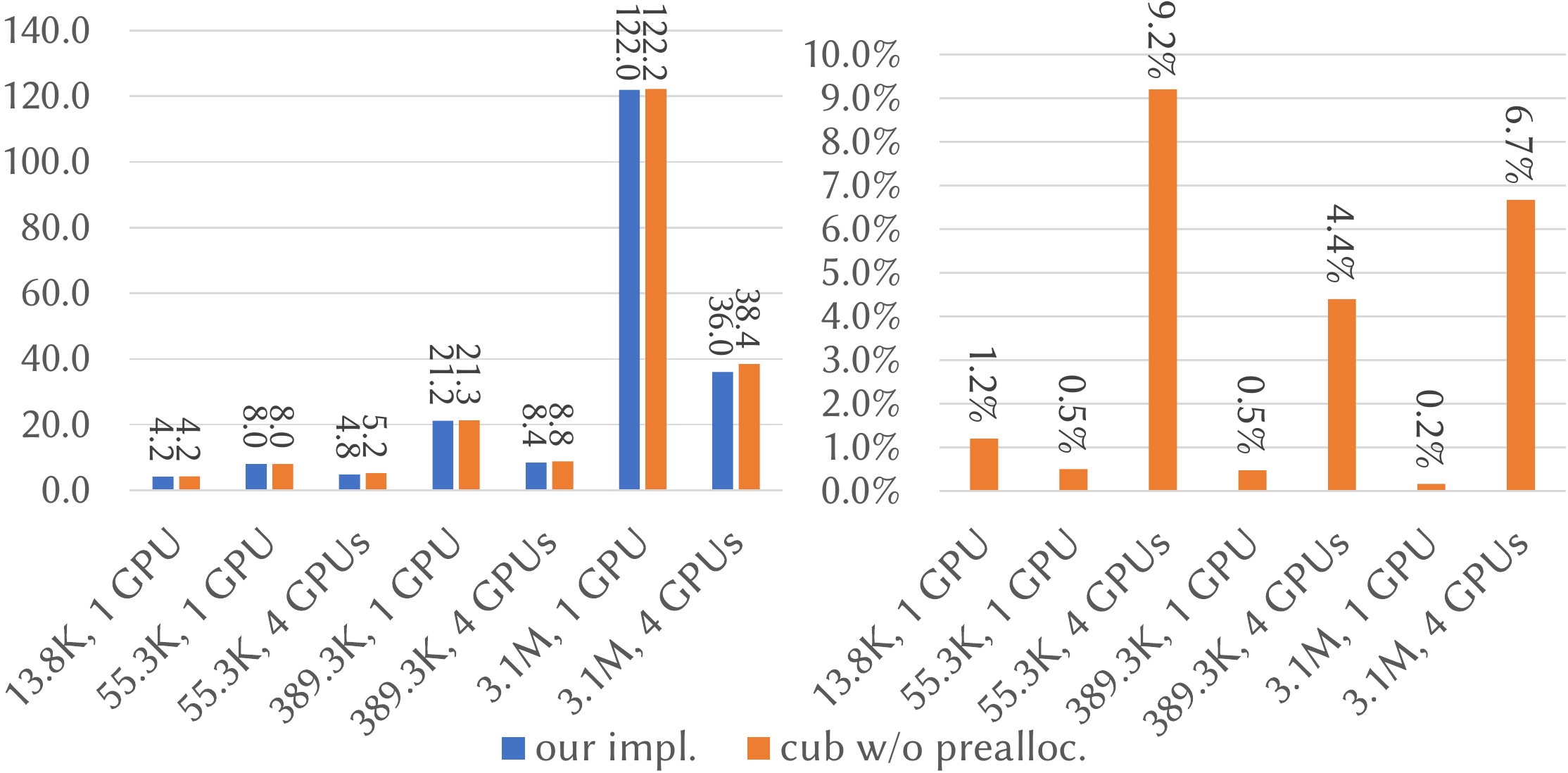}
    \caption{\textbf{Left:} Timing (in milliseconds) comparison between our implementation and the one does not pre-allocate the scratch memory for the CUB library to perform prefix sums; \textbf{Right:} The percentage of additional costs incurred by not pre-allocating the scratch memory.}
    \label{fig:single_gpu_cub_alloc}
\end{figure}
\begin{figure}[htb]
\centering
    \includegraphics[width=1.0\linewidth]{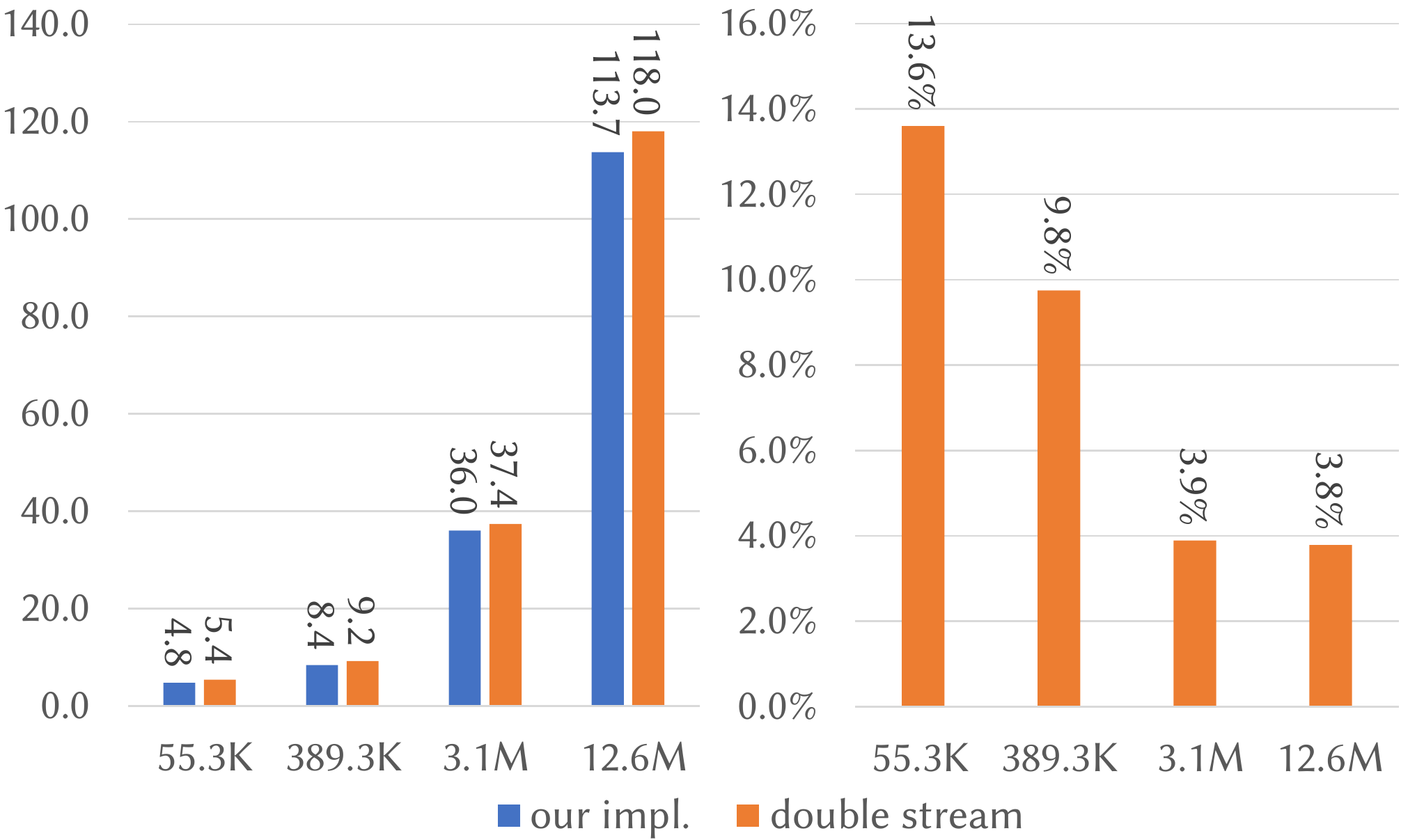}
    \caption{\textbf{Left:} Timing (in milliseconds) comparison between our implementation and the one uses two GPU streams to update the nodal velocities after P2G transfer (one stream for the non-shared particle blocks and another stream for the particle blocks shared between multiple GPUs) to manually overlap the computation with inter-GPU data transfer; \textbf{Right:} The percentage of additional costs incurred by using two streams. This test involves 4 GPUs for all the scenarios.}
    \label{fig:multi_gpu_stream}
\end{figure}
\begin{figure}[htb]
\centering
    \includegraphics[width=1.0\linewidth]{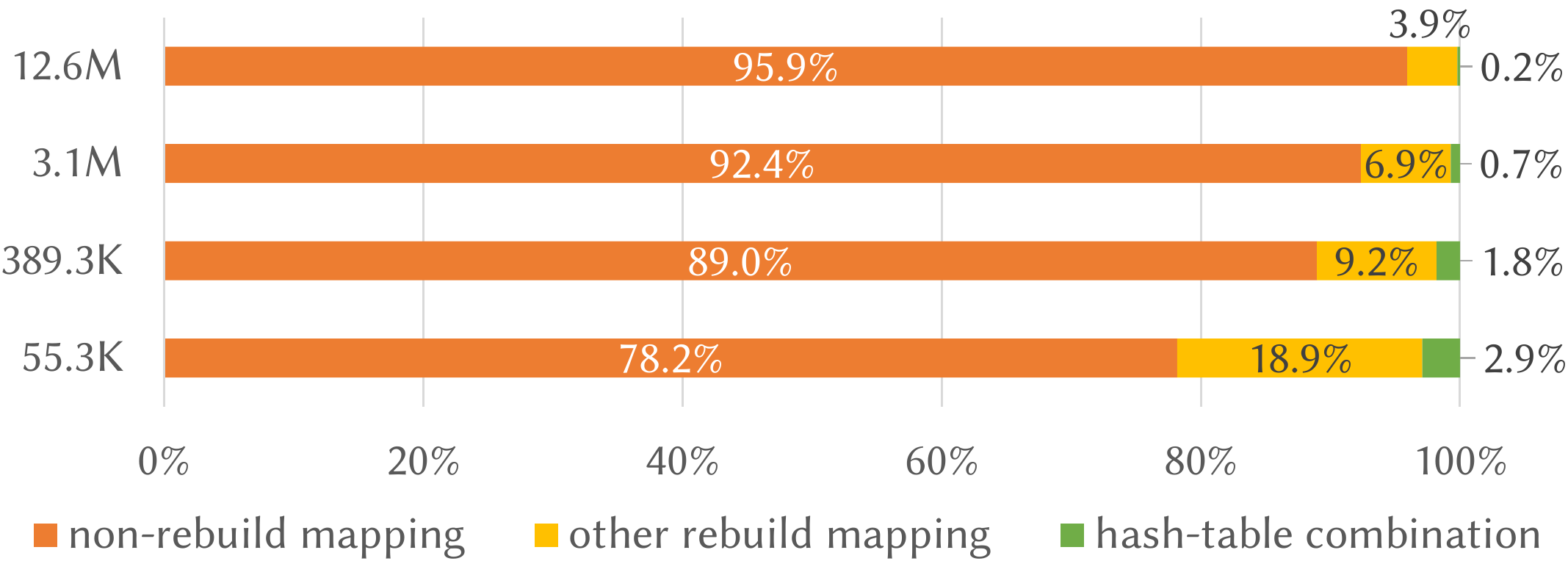}
    \caption{Percentage of time spent for simulating one frame of different size of particles. The \emph{non-rebuild mapping} phase includes P2G and G2P transfer, as well as stress computation during P2G and boundary condition solve during updating the nodal velocities; the \emph{other rebuild mapping} phase includes computing new particle indices and re-sorting particles in order; and the \emph{hash-table combination} phase indicates combining the hashing results from multiple GPUs. This test involves 4 GPUs for all the scenarios.}
    \label{fig:multi_gpu_hashing_percentage}
\end{figure}
\begin{figure}[htb]
\centering
    \includegraphics[width=1.0\linewidth]{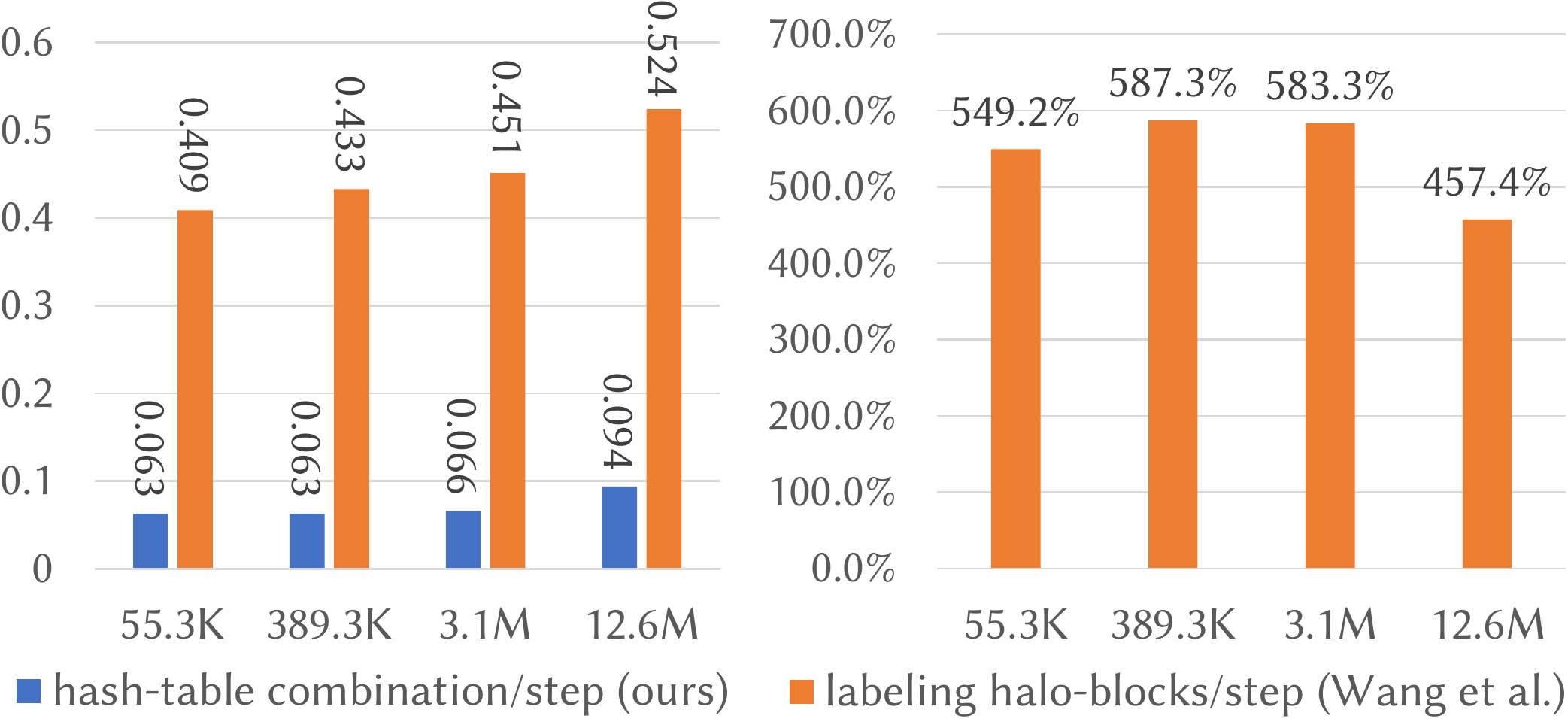}
    \caption{\textbf{Left: \emph{Per step}} timing (in milliseconds) comparison between our lightweight hash-table combination and the halo-blocks' labeling used by Wang et al.~\shortcite{wang2020massively}; \textbf{Right:} The percentage of additional costs incurred by the halo-blocks' labeling, compared with our lightweight hash-table combination. This test involves 4 GPUs for all the scenarios.}
    \label{fig:multi_gpu_hashing_step}
\end{figure}
\begin{figure}[htb]
\centering
    \includegraphics[width=1.0\linewidth]{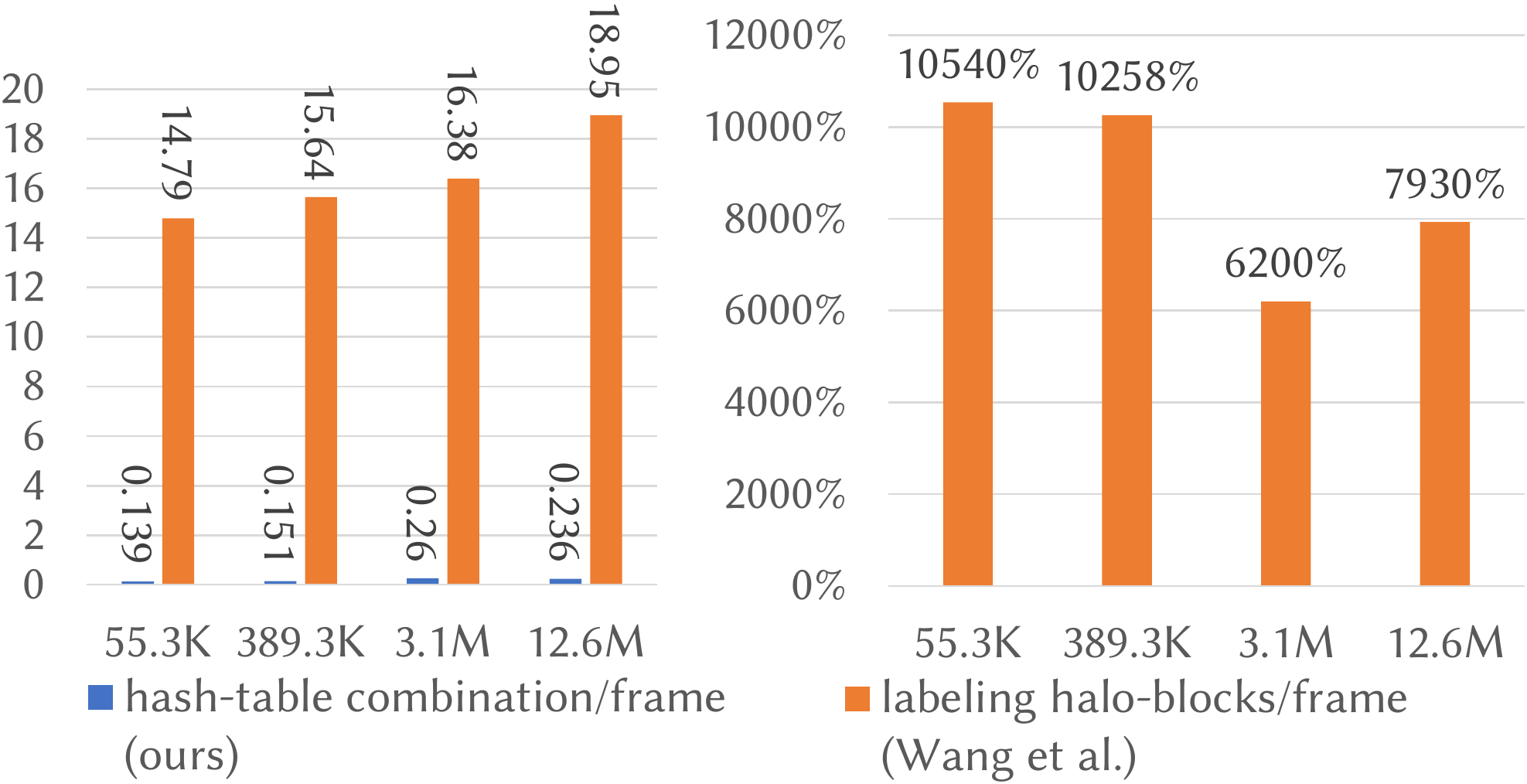}
    \caption{\textbf{Left: \emph{Per frame}} timing (in milliseconds) comparison between our lightweight hash-table combination and the halo-blocks' labeling used by Wang et al.~\shortcite{wang2020massively}; \textbf{Right:} The percentage of additional costs incurred by the halo-blocks' labeling, compared with our lightweight hash-table combination. This test involves 4 GPUs for all the scenarios.}
    \label{fig:multi_gpu_hashing_frame}
\end{figure}

In multi-GPU applications, the efficiency is defined as~\cite{kraus2019multi} 
\begin{equation}
   e = \frac{t_1}{n \times t_n}
\label{eq:efficiency}
\end{equation}
where $t_1$ and $t_n$ are the time consumed with only one GPU and with $n$ GPUs respectively.
In general, we have $e \leq 1$. $e$ only reaches 1 (or $100\%$) in the ideal case, where $n$-GPUs produce $n$-times performance scaling, making $t_n = t_1/n$. The closer $e$ gets to $1$, the better a multi-GPU implementation/algorithm would be.
As the first effort of optimizing MPM on multiple GPUs, Wang et al.~\cite{wang2020massively} build up a working pipeline from scratch and deliver good performance. 
Following the principles proposed in~\secref{pcpmulti}, we adopt several new schemes that successfully push the efficiency of our pipeline closer to the ideal case, as shown in~\figref{compare_wang}.

\subsection{Tag shared and non-shared blocks}
Wang et al.~\shortcite{wang2020massively} tagged a block into three different types, assigned blocks in which particles reside, transferred block to which the particles may write, and advected blocks to which particles may advect. Based on this tagging, they also identify the ``halo regions" containing overlapping blocks on two or more cards. Instead of using such a sophisticated scheme, in this work, we adopt a much simpler scheme that is far less time-consuming.

We tag blocks into two types: blocks shared by other GPUs and blocks not shared by any GPUs. Blocks with the latter type can be processed similarly as blocks on a single GPU, while shared blocks also need a pass of data reduction after P2G transfer. To identify the shared blocks, we maintain a list storing all the block codes on each GPU during the construction of the hash table. An inter-GPU synchronization is enforced right before the tagging to guarantee that the code lists from all GPUs are in place. Then each GPU reads block codes from other GPUs via peer-to-peer (P2P) read operation, computing their spatial indices through block codes. For each incoming block, we then check whether its code already exists in the local hash table: if so, both its source GPU id and memory address are recorded for later reduction.
As we adopt P2P read, on each GPU, our tagging step is entirely independent of all other GPUs, i.e., after the tagging step, each GPU can move forward without a second inter-GPU synchronization that may stall the devices.

Wang et al.~\shortcite{wang2020massively} conducted tagging by copying data through \emph{cudaMemcpy} both between GPU and CPU and between GPUs along with multiple CPU-GPU and inter-GPU synchronizations. On the contrary, our method is relatively lightweight and only requires one inter-GPU synchronization, leading to $5\times$ acceleration per step as shown in \figref{multi_gpu_hashing_step}.
Due to our free-zone scheme (\secref{rbmdetail}), this tagging step is conducted only when rebuild-mapping happens (while transfers among GPUs are still performed every time step), which tremendously reduces the cost per frame as demonstrated in~\figref{multi_gpu_hashing_frame}.

\subsection{Compute the reduce sum on shared blocks}
For each time step, as the values on the nodes of shared blocks are distributed to different cards, a reduction operation has to be applied across multi-GPUs to get the correct nodal values.
By following the multi-GPU principles, we propose a method that is both lightweight and quite efficient.

For collecting nodal values stored on shared blocks from different GPUs, we synchronize all GPUs right after the P2G step as peer-to-peer read operation will be used in the reduction step. This operation is the only inter-GPU synchronization required in each step if the rebuild mapping is not initiated.
We further fuse the reduction with the original nodal computations (i.e., updating nodal velocities and resolving collisions) into a single kernel to overlap the inter-GPU communications and local computations through warp scheduling, which provides better performance (refer to \secref{multi_gpu_stream} for statistical details).

To ensure each GPU proceeds independently and avoid data race, we adopt a double-buffering strategy to isolate the local reduction on each GPU.
Specifically, we allocate two buffers for the P2G transfer and inter-GPU reduction. We store the results from each GPU's P2G on the second buffer. 
For each shared block, we then read the copies of this block from other GPUs' second buffer via P2P-reads, add them together with the local copy, and store the sum to the first buffer of the current GPU. In this process, we keep each GPU's P2G results (stored in the second buffer) untouched so that any other card would be able to read its original value via P2P reads.

\paragraph{Remark.} 
Our method effectively conducts multiple reductions for the same shared pblock. An alternative is to compute and store on a single GPU while directly writing the values to other GPUs via P2P atomics, which removes the need for an additional buffer. However, inter-GPU atomics with a large portion of data may need many (implicit) synchronizations between GPUs~\cite{abe2014gdev}, which can be very ineffective (\secref{multi_gpu_sbaiw}).

Similar to the case of a single GPU, as our free zone scheme introduces more blocks than required for each single time step, we may have more P2P transfers than needed. To alleviate this issue, we skip the transfers from the marked blocks that are not touched by particles on the source GPU in the current time step.

\begin{figure*}[htb]
\centering
    \includegraphics[width=1.0\linewidth]{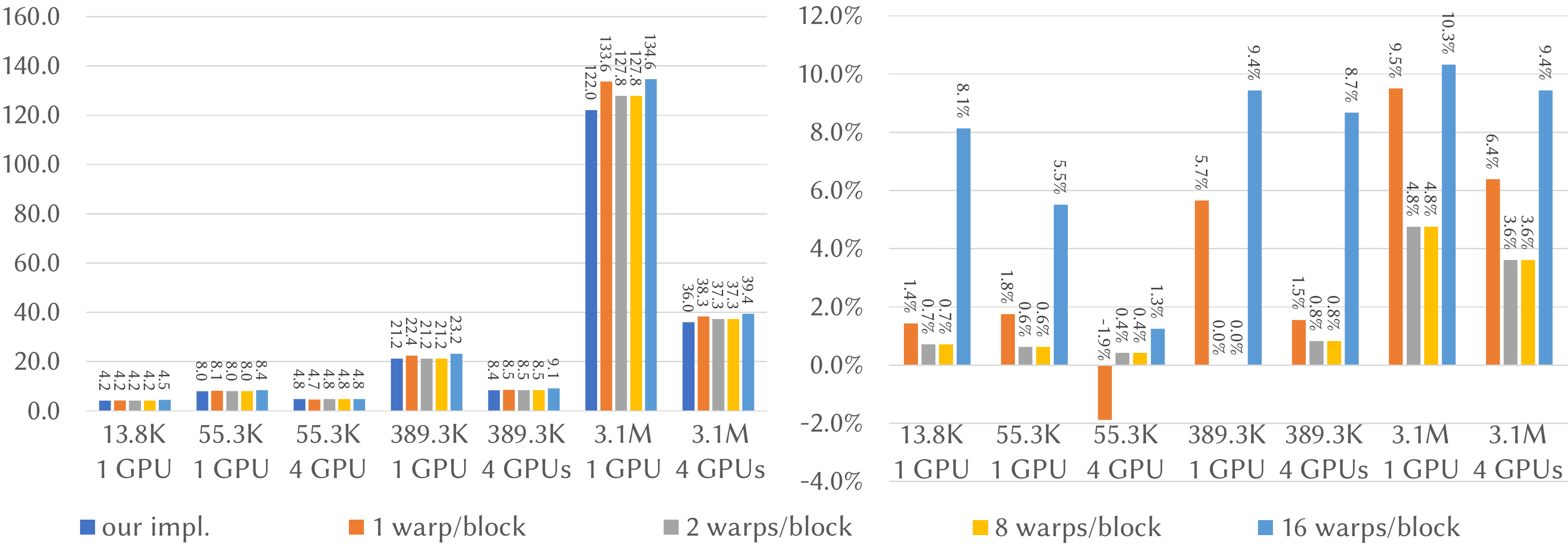}
    \caption{\textbf{Left:} Timing (in milliseconds) comparison between implementations with a different number of warps used for each CUDA block (our implementation uses four warps per block); \textbf{Right:} The percentage of additional costs incurred by different designs other than ours. When the number of particles is tiny, e.g., 13.8K on one GPU (or 55.3K on four GPUs), it is not ensured that four warps per block are always optimal. Using one warp per block may increase the granularity and reduce the tail effect. Nevertheless, we adopt four warps per block since this choice performs better than the others in most cases.}
    \label{fig:single_gpu_warp_num}
\end{figure*}
\section{Results\label{sec:results}}

Following principles proposed in~\secref{principles}, we make substantial modifications in an explicit MLS-MPM~\cite{hu2018moving} framework implemented based upon the work by Gao et al.~\shortcite{gao2018gpu} even though some of them may seem to be controversial.
While the final pipeline delivers quite a massive boost in performance compared to state-of-the-art~\cite{wang2020massively}\footnote{We adapt from the open-source code that the authors released on GitHub, \url{https://github.com/penn-graphics-research/claymore}, commit 4a6b0cf on Nov. 21, 2020, adding our benchmark scenario and the neo-Hookean sand model from Yue et al.~\shortcite{yue2018hybrid}.}, a thorough ablation study would help researchers understand the effect of every single modification.

The machine we use to run the benchmarks is an NVIDIA DGX-1. While there are eight Tesla V100 GPUs in total, they are divided into two groups, and only the GPUs in the same group have fully connected NVLinks. As a result, we test up to four GPUs in the same group.
The DGX-1 has two Intel Xeon Platinum 8255C CPUs running at 2.50GHz, and each of them has 24 logical cores and 48 threads. 
The operating system is CentOS 7, the GPU driver version is 460.32.03, and the CUDA version is 11.2. We turn on the compiler option \texttt{-use\_fast\_math} to use native intrinsics by default.

\begin{figure}[htb]
\centering
    \includegraphics[width=1.0\linewidth]{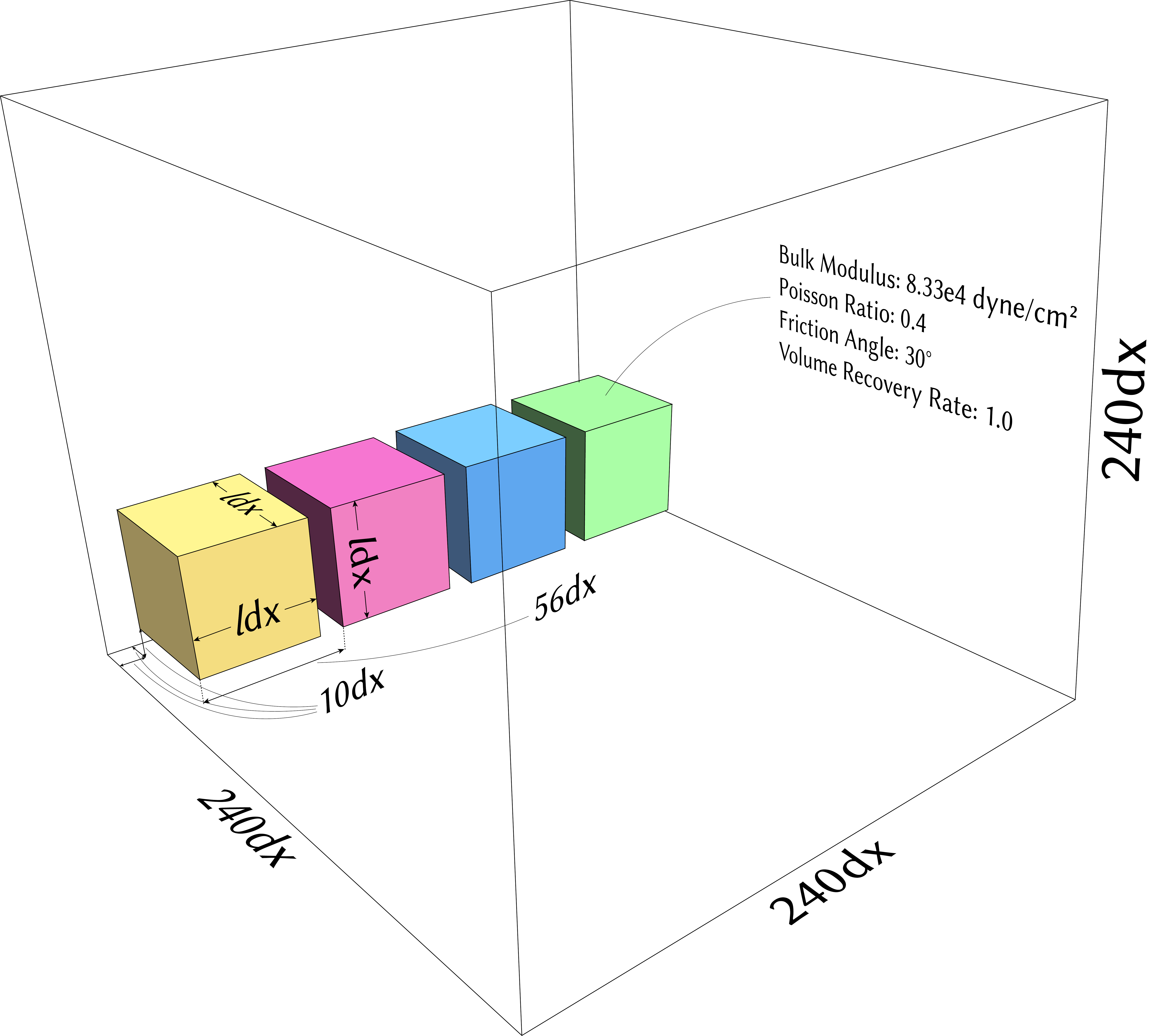}
    \caption{\textbf{The setting of \emph{Sand Blocks}.} We take the case with four boxes as an example. The size of boxes changes in different cases, where we use $l=12$, $l=23$, and $l=46$ for cases that have 55.3K, 389.3K, and 3.1M (or 12.6M) particles, respectively. We use cell size $\text{dx}=25/64\text{cm}$, and within each cell, we initially sampled eight particles in a stratified random way. We adopt Yue et al.~\shortcite{yue2018hybrid}'s neo-Hookean sand model for all the cases.}
    \label{fig:test_scene_setting}
\end{figure}
We begin our profilings with a \emph{highly dynamic} scenario containing several \emph{Sand Blocks} (\figref{test_scene_setting}). With the cell size being fixed, one, four, or sixteen cubic boxes of sand particles are dropped into a container so that the particles collide with the frictionless boundary and with each other to generate energetic dynamics. 
We simulate 60 frames, and each frame has a duration of $1/48$ seconds. We fix the time step size to $5.787\times10^{-4}$ seconds such that each frame consists of 36 time-steps.
We report the average cost per frame across the entire 60 frames.
For the multi-GPU profilings in this scenario, we compute the longest axis of the particle cluster and sort the particles along with this axis. Then we divide the particles by their indices and distribute them evenly onto four GPUs.

\subsection{Compare with state-of-the-art}
Our final pipeline is compared with the one proposed in Wang et al.~\shortcite{wang2020massively}. 
As shown in~\figref{compare_wang} (bottom), their implementation consumes much more time than ours, ranging from $96\%$ to $1385\%$.
Our pipeline is exceptionally faster than Wang et al.'s when the number of particles on each GPU is small.

We also compare the four-GPU efficiencies (the capability of scaling) between our final pipeline and the one proposed in Wang et al.~\shortcite{wang2020massively}. We compute the efficiency $e$ through equation (\ref{eq:efficiency}) from the averaged timings. Our multi-GPU pipeline almost reaches the ideal case (i.e., $e\approx 1$) when the number of particles is large ($96.3\%$ for 12.6M) and keeps an acceptable efficiency when the number of particles is small ($38.8\%$ for 55.3K). On the contrary, Wang et al.'s efficiencies drop to less than $25\%$ when the simulation contains less than 389.3K particles, indicating too much overhead in their multi-GPU pipeline, making their four-GPU simulation even slower than their 1-GPU simulation.

\paragraph{Remark:} In this subsection, we adopt G2P2G when the number of particles per GPU is less than 100K (with either multiple GPUs or a single GPU) as the instruction miss rate is still tolerable.

\subsection{Ablation study on a single GPU\label{sec:ablation_single}}
\begin{figure}[htb]
\centering
    \includegraphics[width=1.0\linewidth]{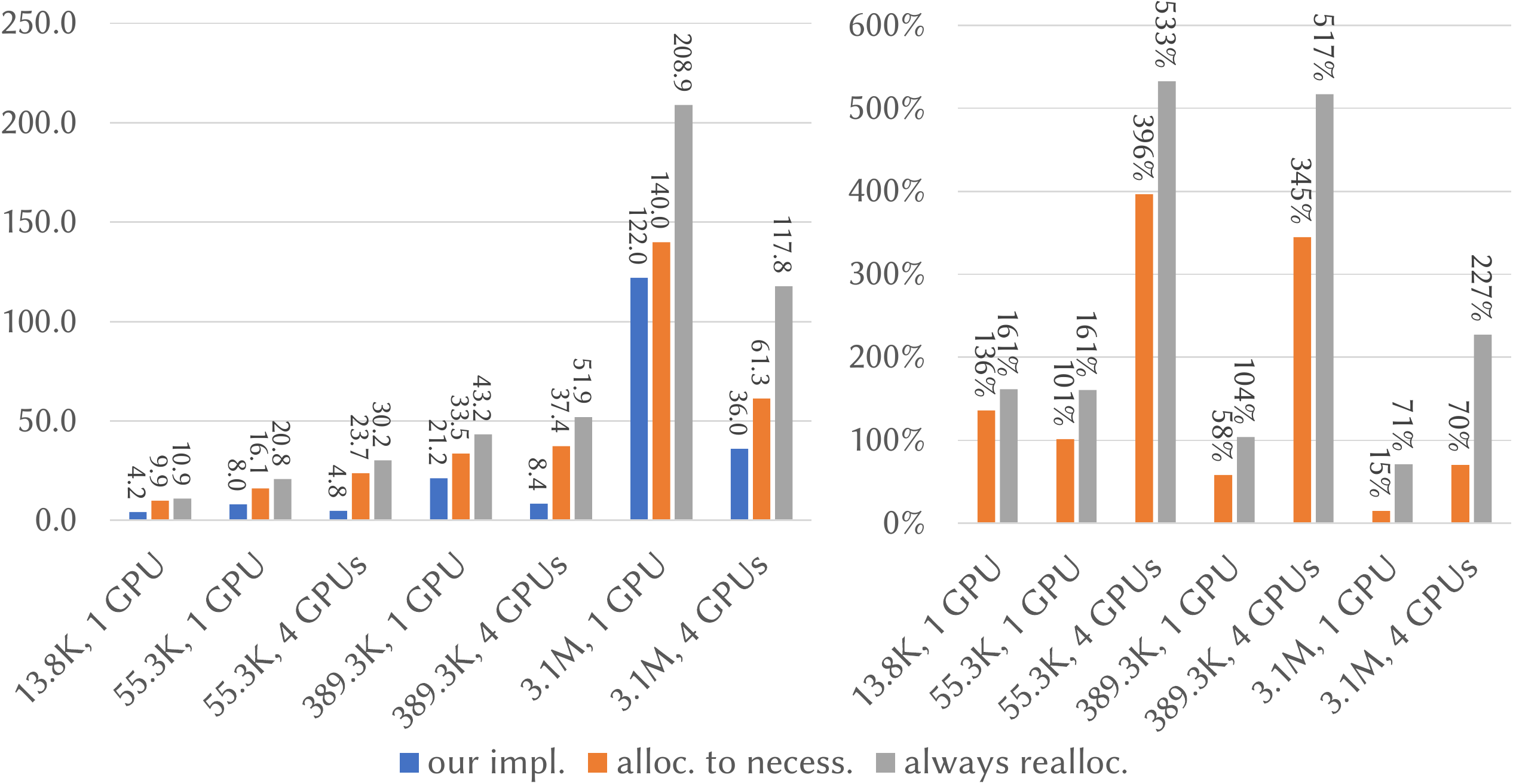}
    \caption{\textbf{Left:} Timing (in milliseconds) comparison between our memory allocation scheme, allocating to a size necessary at current step (\emph{alloc. to necess.}), and always re-allocating regardless of sizes (\emph{always realloc}); \textbf{Right:} The percentage of additional costs incurred by different schemes other than ours.}
    \label{fig:memory_realloc}
\end{figure}
In the following sections, we perform an ablation study on our final pipeline. We roll back and remove a single scheme derived from our principles and compare its performance with our final pipeline. We stick to the traditional scheme where P2G and G2P are two separate steps for the ease of experiment.
\subsubsection{Reduce memory reallocation\label{sec:reduce_alloc_single}}
In~\figref{memory_realloc} we show that our memory allocation scheme effectively reduces the need for reallocation and dramatically eliminates overhead. We compare with other schemes that are commonly adopted. For example, suppose we only allocate to the required size (in other words, we never allocate more than necessary). In that case, the overhead is $15\%$--$136\%$ for simulation on a single GPU. On the other hand, if we always reallocate by each step, regardless of the allocation size, the overhead would increase up to $161\%$ on a single GPU. More discussion about memory reallocation for multiple GPUs can be found in~\secref{reduce_alloc_multi}.

\subsubsection{Minimize rebuilding mappings} 
Our new scheme only rebuilds the mapping between particles and the background grid every $10$--$30$ time steps (across all the test cases' frames). 
When running with multiple GPUs, a standard rebuild-mapping step consists of more than ten CUDA kernels, CPU-GPU synchronizations, and inter-GPU synchronizations. 
In~\figref{single_gpu_rbm}, we show that without our scheme, i.e., rebuilding the mapping every time step, the computational cost would increase by $42.5\%$--$266.1\%$ per frame. 

\subsubsection{Keep particles sorted with cheaper sorting\label{sec:result_sp}}
\begin{figure}[htb]
\centering
    \includegraphics[width=1.0\linewidth]{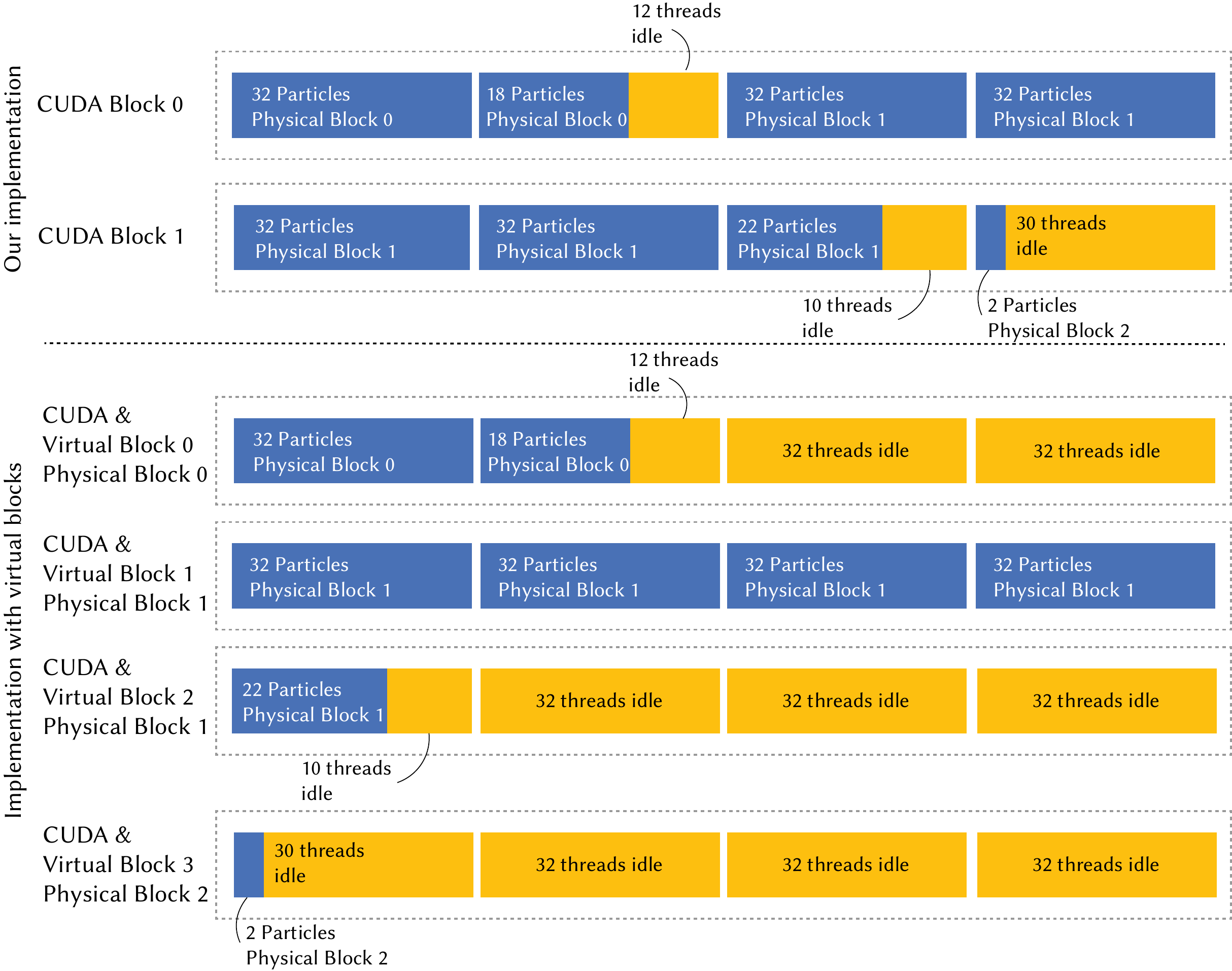}
    \caption{\textbf{Comparison between assigning a CUDA block to particles in mixed physical blocks (\emph{top}) and assigning to a unique virtual block, each of which only holds particles from a single physical block (\emph{bottom}).} The latter scheme, proposed by Gao et al.~\shortcite{gao2018gpu}, may have more CUDA blocks being launched and less efficient utilization of threads when the size of physical blocks varies. Here we have three physical blocks holding 50, 150, and two particles, respectively. A solid box indicates a CUDA warp with 32 threads, and a dashed box indicates a CUDA block with four warps.}
    \label{fig:vblock_diagram}
\end{figure}
In our pipeline, a complete sort-particle step, i.e., sorting particles to cells, only occurs with the rebuild-mapping step. A warp-level radix sorting is used to help reduce the number of atomics in other time steps.
In~\figref{single_gpu_sorting} we compare with two alternatives. 
The first one gets rid of our warp-level radix-sort, leaving particles unsorted (\emph{P2G w/o sorting}) for the time steps between two rebuild-mappings.
The cost per frame increases by $1.4\%$--$16.2\%$ compared to ours.
The second option conducts a complete sort-particle step every time step and introduces $11.3\%$--$24.8\%$ additional cost (\emph{P2G w. full sorting}).

\subsubsection{Avoid reduction on shared memory}
Two kinds of methods have been designed to handle the write conflicts in the P2G steps. Gao et al.~\shortcite{gao2018gpu} adopt three levels of reductions: first, a warp-level reduction reduces the number of atomics within each warp; second, each warp writes the sums into the shared memory via shared atomics; and a \emph{syncthreads} function synchronizes the threads in the CUDA block before the final reduction is applied to write to the global memory.
Wang et al.~\shortcite{wang2020massively} and Hu et al.~\shortcite{hu2019taichi} avoid the first reduction, and they shuffle particles such that there would be minimum conflicts within each warp.
On the contrary, we eliminate the second reduction, which relies on the non-native shared atomics of floating-point numbers, to follow the proposed principle. In other words, after the warp-level reduction, we directly write to the global memory via global atomics without using the shared memory as the scratchpad.
In~\figref{single_gpu_vblock}, we conduct extensive comparisons between different schemes to show that our scheme manifests the best efficiency.

Compared with our scheme, partitioning particles into virtual blocks not only needs extra computations ($6.4\%$--$27.0\%$ additional cost as in \figref{single_gpu_vblock}, \emph{v.block}), but also introduces workload imbalance (\figref{vblock_diagram}) as virtual blocks would have many more idle threads. Notice that virtual blocks from the same particle block will write to the same global memory addresses, potentially increasing the number of conflicts.

Reducing to the shared memory also requires an additional in-block synchronization which turns out to be quite expensive. 
Putting an extra synchronization at the end of our implementation of P2G, we can observe (through Nsight Compute) the amount of warp stall barriers doubles. Such barriers prevent the warps from switching to handle another block of particles. 
As a consequence, the cost dramatically increases by $48.0\%$--$120.6\%$ per frame (\figref{single_gpu_vblock}, \emph{v.block w. final sync}).

Furthermore, there do not exist native GPU instructions that support atomic addition for floating-point numbers on shared memory in most recent NVIDIA graphics cards. 
The compiler would translate the shared atomics into loops and atomic compare-and-swap as shown in~\figref{atomic_disassembly}. Such a translation would increase the number of instructions by $38\%$ and impact the computational performance. \figref{single_gpu_vblock}, \emph{v. block w. shared atomics}, shows there is $131.1\%$--$247.1\%$ additional cost when using both virtual blocks and shared atomics.

Finally, as the size of the shared memory in each CUDA block is relatively large in MPM, the GPU occupancy becomes much lower than our implementation. Increasing the number of threads per block may help increase the occupancy but would not improve the performance. 
When the virtual block has a larger size, more threads stay idle for the particle blocks that do not have enough particles. \figref{single_gpu_vblock}, \emph{v. block w. sh. a. + occu.}, shows the situation deteriorates when increasing the number of threads per block from 128 to 256.

\subsubsection{Relieve register pressure\label{sec:relieve_registers}}
\begin{figure}[htb]
\centering
    \includegraphics[width=1.0\linewidth]{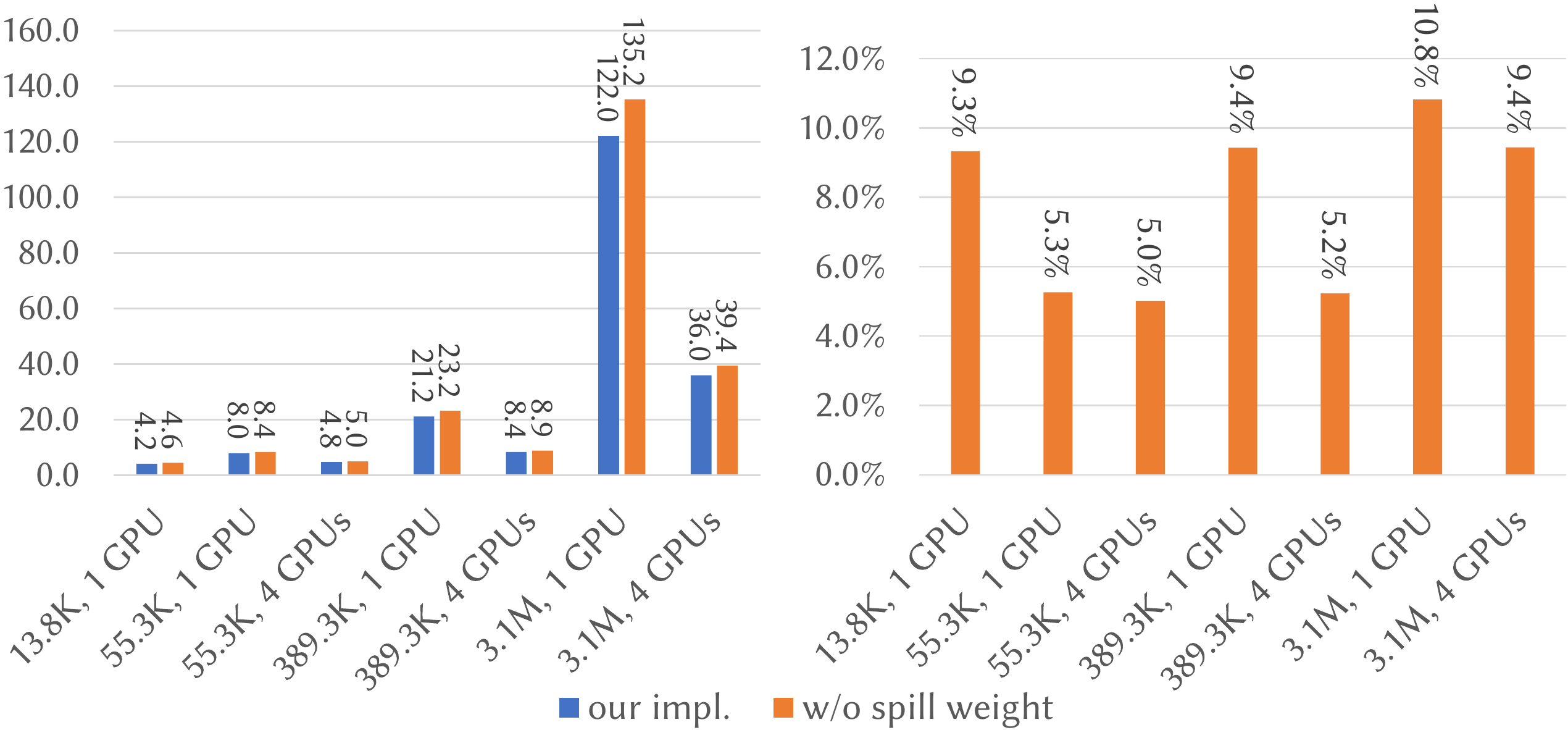}
    \caption{\textbf{Left:} Timing (in milliseconds) comparison between our implementation and the one without spilling B-spline weight to the shared memory; \textbf{Right:} The percentage of additional costs incurred by not spilling.}
    \label{fig:single_gpu_spill}
\end{figure}
Merging small kernels into large ones also means these large kernels consume a lot more registers, resulting in either register spilling or low occupancy. Instead of storing temporary results in registers, we can also store them in shared memory to relieve register pressure.
In~\figref{single_gpu_spill}, we compare the performance of storing the B-spline weights in particle-grid transfer steps to either the shared memory or to the registers, and we show that the latter would bring in $5.0\%$--$10.8\%$ more costs as the occupancy gets decreased.

\subsubsection{Use AoSoA instead of SoA}
We adopt the Array-of-Structs-of-Array (AoSoA) data structure proposed in Wang et al.~\shortcite{wang2020massively} for storing properties on both the particles and the background grid.
We compare AoSoA with the Structs-of-Array (SoA) data structure used in Gao et al.~\shortcite{gao2018gpu}.
As shown in~\figref{single_gpu_storage}, the SoA storage would introduce $6.7\%$--$11.0\%$ additional cost per frame.

Memory requests in modern GPUs are fulfilled in fixed width (i.e., cacheline~\cite{nvidia2021kernel}), and when requested data is not aligned with this width, more access operations are enforced to complete a request.
For example, when reading some data with the exact size of a cacheline, if this chunk of data is misaligned with the cacheline, two read operations are required instead of one, leading to less efficiency.
AoSoA packs data into small $32\times\sharp\text{channels}$ packets, and thus the memory access pattern of AoSoA is well aligned to the cacheline. 
In contrast, SoA has less aligned requests and needs $8\%$--$18\%$ more sectors (and thus more data) sent/received than AoSoA.

Furthermore, at the end of P2G, when threads in the same warp write to different grid nodes, SoA presents a less coalesced pattern, i.e., accessing addresses within the same warp can be wildly different from AoSoA, leading to $16\%$ more accesses.

\subsubsection{Merge small kernels\label{sec:mkarmo}}
\begin{figure}[htb]
\centering
    \includegraphics[width=1.0\linewidth]{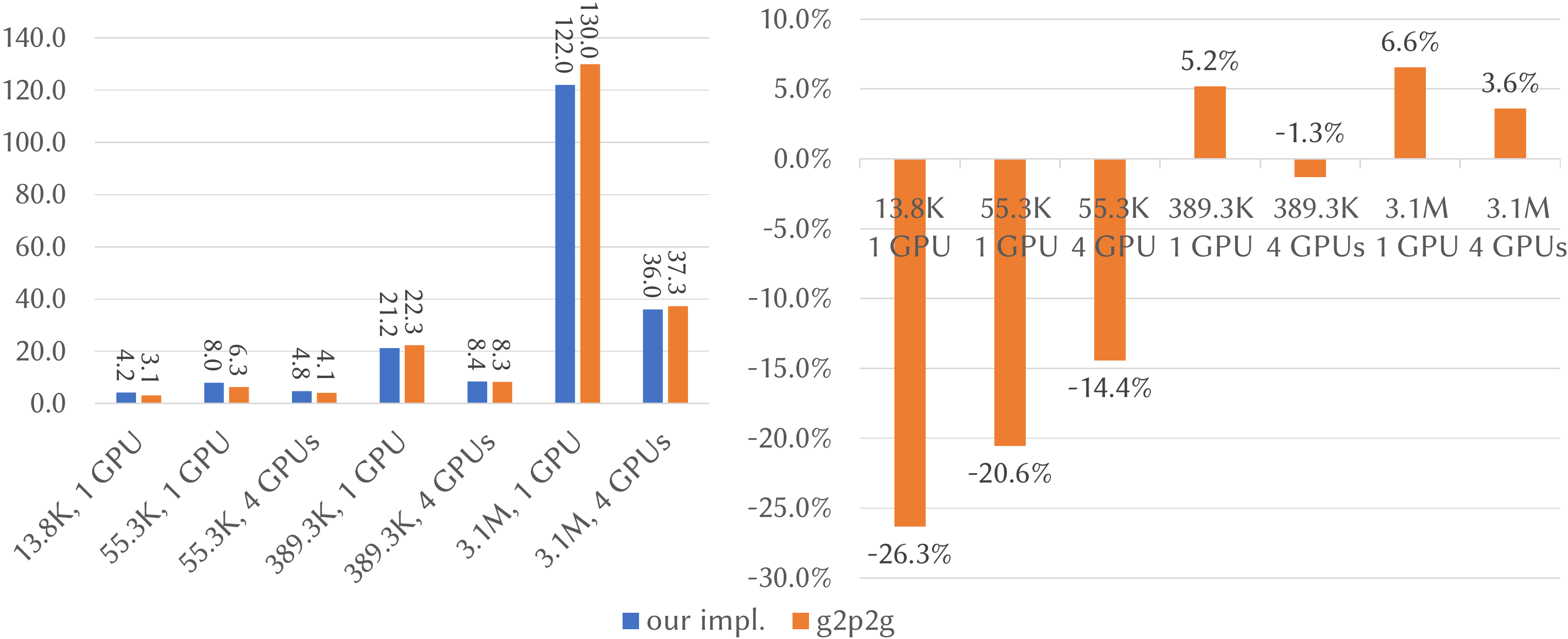}
    \caption{\textbf{Left:} Timing (in milliseconds) comparison between our implementations with and without G2P2G used; \textbf{Right:} The percentage of additional costs (or decreased costs if negative) incurred by different designs other than the default version (without G2P2G).}
    \label{fig:single_gpu_g2p2g}
\end{figure}
In~\figref{single_gpu_indiv}, we show that merging small kernels into appropriate size boosts the performance as certain global memory operations, e.g., writing and then reading intermediate variables, can be removed.
Compared with our final pipeline, computing stress with an individual kernel before P2G transfer (\emph{indiv. comp. stress}) induces $4.7\%$--$11.7\%$ additional cost per frame; resolving boundary condition separately after updating nodal velocities (\emph{indiv. resolv. b. c.}) induces $0.2\%$--$4.4\%$ additional cost per frame; and clearing nodal buffers with an individual kernel before P2G transfer (\emph{indiv. buffer. clearing}) brings in as much as $4.3\%$ additional cost per frame. 
While the small kernels cannot fully utilize the SM (also known as the \emph{tail effect}~\cite{micikevicius2012gpu}), the SM has to finish execution of these kernels before executing any other larger kernels afterward. By integrating these small kernels into the larger kernels following them, the limited computation or memory operations can be overlapped with other heavier computations or memory operations through interleaved warp scheduling, reducing the tail effect with small kernels.

In~\figref{single_gpu_g2p2g} we show that merging P2G and G2P transfer into a single G2P2G transfer, as suggested by Wang et al.~\shortcite{wang2020massively}, is only beneficial when the number of particles is small as the instruction cache miss rate is still tolerable.

\subsubsection{Tune the size of CUDA blocks}
Our pipeline divide particles into warps, and each CUDA block handles four warps.
In~\figref{single_gpu_warp_num}, we compare the performance with a different number of warps per CUDA block.
The number of simultaneously active blocks on one SM is limited.
Thus, when the number of particles is extensive, using 1 or 2 warps per CUDA block would lead to a large number of blocks, and some of them have to stay inactive.
On the other hand, CUDA blocks with many warps would need too many registers, resulting in low occupancy.

\subsubsection{Adopt fast-math intrinsics\label{sec:result_fast_math}}
\begin{figure}[htb]
\centering
    \includegraphics[width=1.0\linewidth]{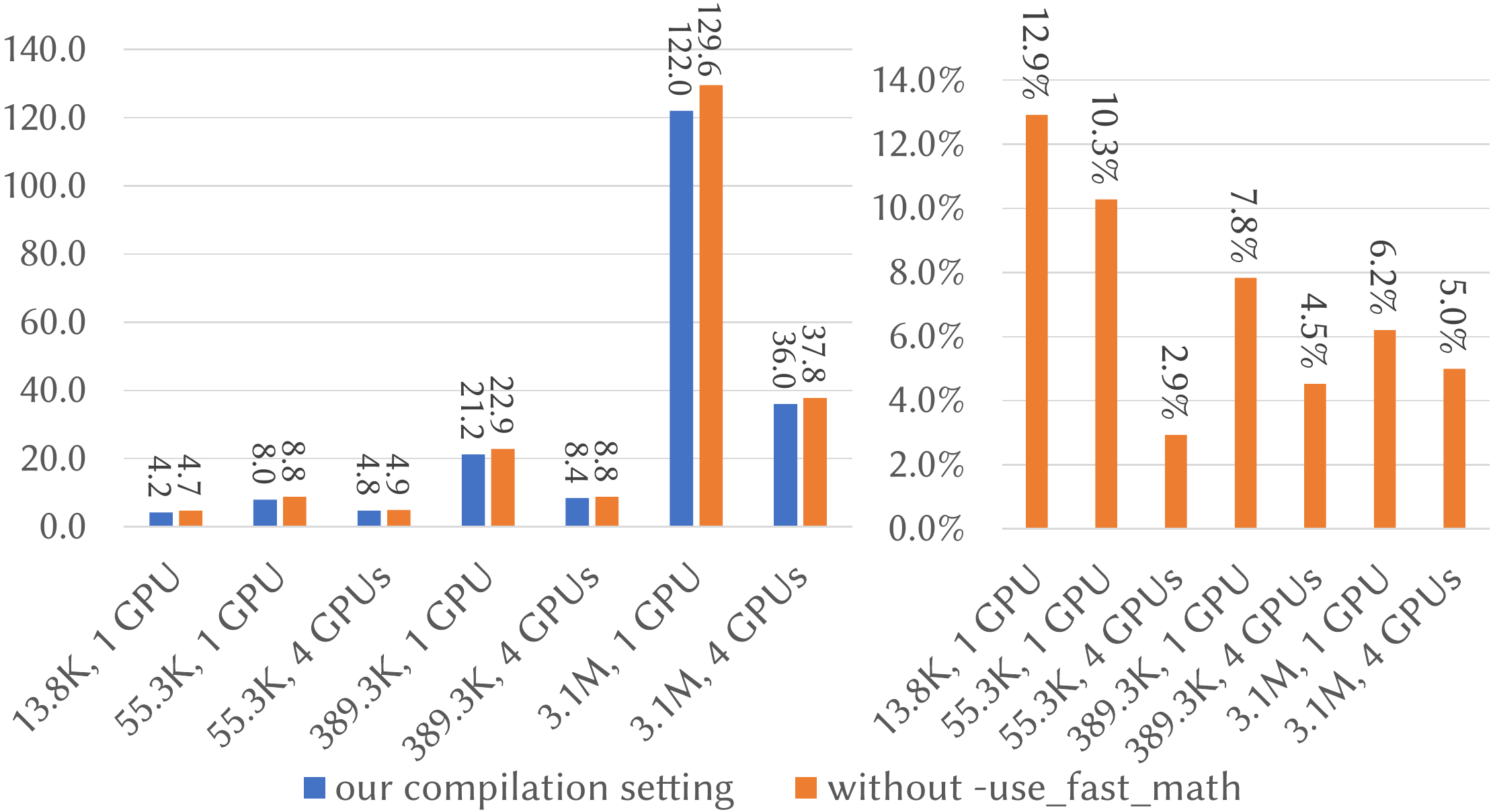}
    \caption{\textbf{Left:} Timing (in milliseconds) comparison between the binary compiled with \texttt{-use\_fast\_math} (as our default setting) and the one without; \textbf{Right:} The percentage of additional costs incurred when disabling fast-math intrinsics.}
    \label{fig:use_fast_math}
\end{figure}
As shown in~\figref{use_fast_math}, disabling the fast-math intrinsics would introduce $2.9\%$--$12.9\%$ extra cost.

\subsection{Ablation study on multiple GPUs\label{sec:ablation_multi}}
\subsubsection{Reduce memory reallocation\label{sec:reduce_alloc_multi}}
The principle for reducing memory reallocation proposed in~\secref{reduce_alloc_single} also applies to multi-GPU environment. 
However, in contrast to single-GPU scenarios, for multi-GPU, when one card reallocates the memory, it also needs to update information with other GPUs and thus inter-GPU synchronizations are enforced. As a result, we can observe much more overhead with frequent memory reallocation in a multi-GPU environment.
As shown in~\figref{memory_realloc}, when simulating with four GPUs without pre-allocating more memory than requested, the overhead can be up to $396\%$.
If we trivially reallocate every time step, the overhead is up to $533\%$. 

One less obvious case is that libraries such as Thrust~\cite{bell2012thrust} and CUB~\cite{merrill2015cub} that provide implementations of parallel algorithms need scratch memory to store the intermediate results. 
Implicitly re-allocating the scratch memory each time these algorithms execute impacts the performance and in~\figref{single_gpu_cub_alloc} we show that it introduces $4.4\%$--$9.2\%$ additional cost when simulating with four GPUs.

\subsubsection{Overlap computation and inter-GPU data transfer with interleaved warp scheduling \label{sec:multi_gpu_stream}}
In our pipeline, the inter-GPU communications and the computations on the grid nodes are merged into one kernel, and thus in one stream. And we rely on the interleaved warp scheduler to automatically overlap the memory accesses and the computations.
The alternative is to overlap the two of them with two different streams manually. 
In~\figref{multi_gpu_stream}, we show that the two-stream implementation induces $3.8\%$--$13.6\%$ additional cost.
As revealed in the analysis of Nsight System, the extra cost comes from the synchronization between the two different streams, while there is almost no difference in performance between the two schemes.

\subsubsection{Lightweight block tagging}
In~\figref{multi_gpu_hashing_percentage} we show our tagging scheme is lightweight and occupies only up to $2.9\%$ of the overall cost per frame even when the number of particles is small ($55.3$K). 
In contrast, the prior work~\cite{wang2020massively} consumes much more time \textbf{per step} in tagging (\figref{multi_gpu_hashing_step}). 
Furthermore, with our free zone scheme, both the rebuild-mapping and the tagging only occur every 10--30 steps, leading to tremendous performance gain \textbf{per frame} (\figref{multi_gpu_hashing_frame}).

\subsubsection{Isolate P2P reads and local reduce sum with double-buffering \label{sec:multi_gpu_sbaiw}}
\begin{figure}[htb]
\centering
    \includegraphics[width=1.0\linewidth]{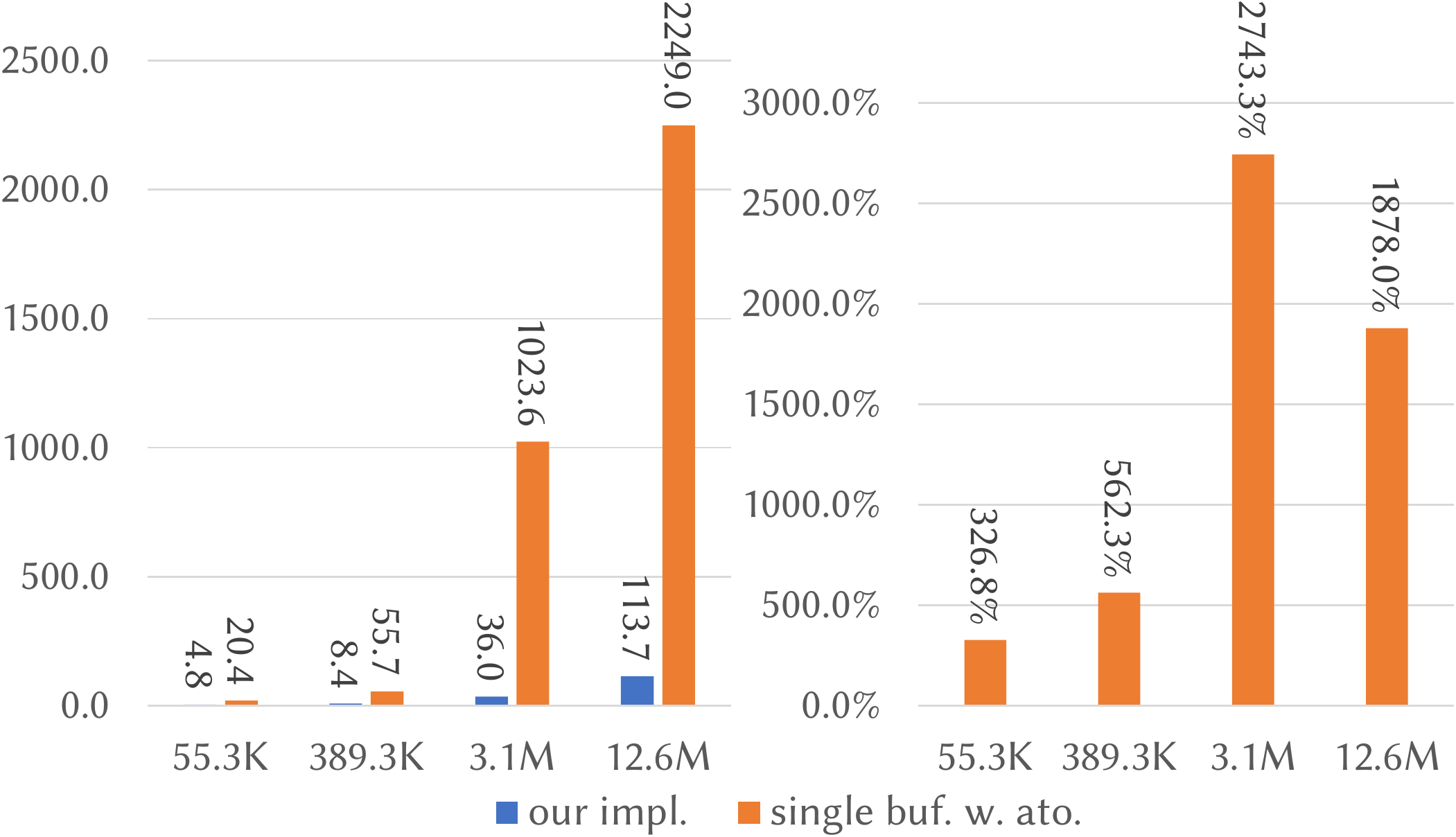}
    \caption{\textbf{Left:} Timing (in milliseconds) comparison between our implementation and the one uses only a single buffer and writes nodal data directly through inter-GPU atomics; \textbf{Right:} The percentage of additional costs incurred with the latter scheme. Four GPUs are used in all scenarios.}
    \label{fig:multi_gpu_sbaiw}
\end{figure}
After the P2G step, each card needs to read values on the shared grid blocks from other cards and then applies a local reduce operation to get the sums.
We adopt a double-buffering scheme such that within one card, the P2P read requests from other cards and the local reduce sum are isolated.
A simple alternative is to complete the reduction via P2P atomics directly.
While it only requires one buffer and thus consumes less memory, this method turns out to be much less efficient than ours as P2P atomics are much more costly than P2P reads and writes, bringing $326.5\%$--$2743.2\%$ additional cost as shown in~\figref{multi_gpu_sbaiw}.

\subsubsection{Inter-GPU barriers}
In~\figref{multi_gpu_mutex} our spinlock scheme is compared with a more standard one that uses the CUDA API \emph{cudaStreamWaitEvent} for an inter-GPU barrier.
In cases with a small number of particles (e.g. $13.8$K per card in our test), \emph{cudaStreamWaitEvent} causes latency on the CPU that consumes $\sim 1.38\text{ms}$ per frame, while the P2G transfer only takes $0.63\text{ms}$. 
We also observe \emph{cudaStreamWaitEvent} costs in $\mathcal{O}(m^2)$ where $m$ is the number of GPUs, which indicates the delay can be more severe for computation with eight cards.

\begin{figure*}[t]
\centering
    \includegraphics[width=1.0\linewidth]{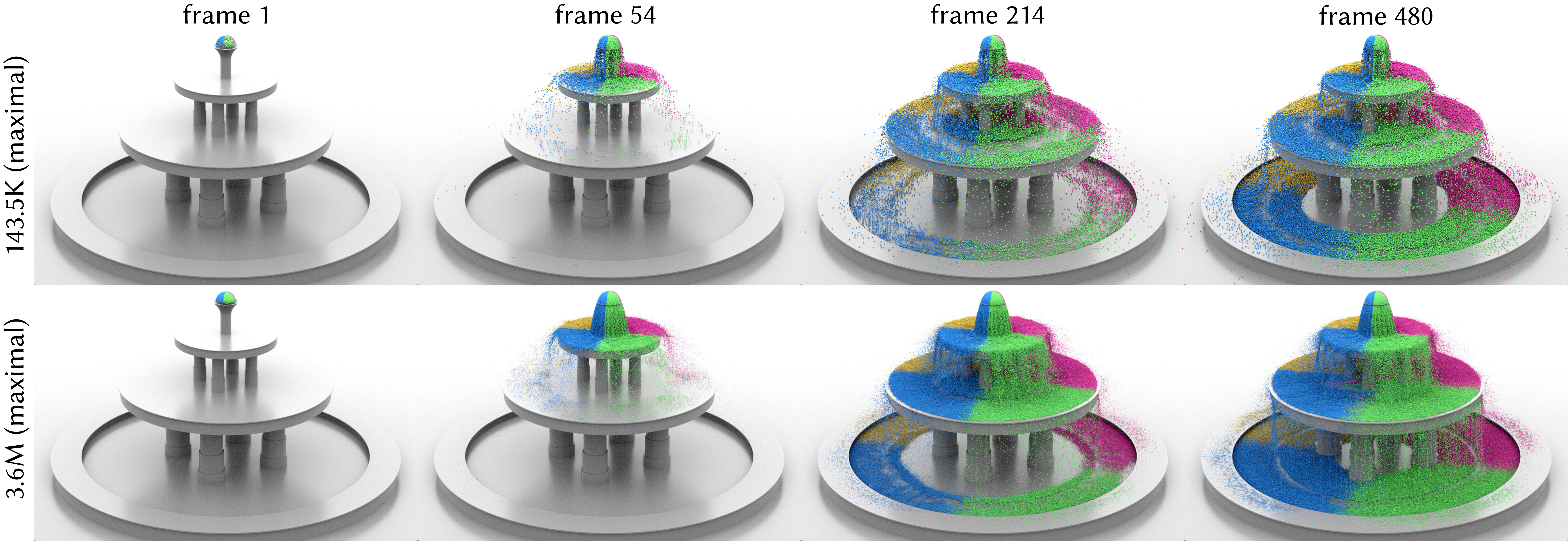}
    \caption{\textbf{Fountain.} Water sprays from the top of a fountain. We show two simulations in different scales. The top one can be simulated in real-time (55.9 frames per second) with four Tesla V100 GPUs, and the bottom is another offline simulation. Particles simulated with different GPUs are colored differently.}
    \label{fig:fountain_rendering}
\end{figure*}

\subsection{Fountain}
\begin{figure}[htb]
\centering
    \includegraphics[width=1.0\linewidth]{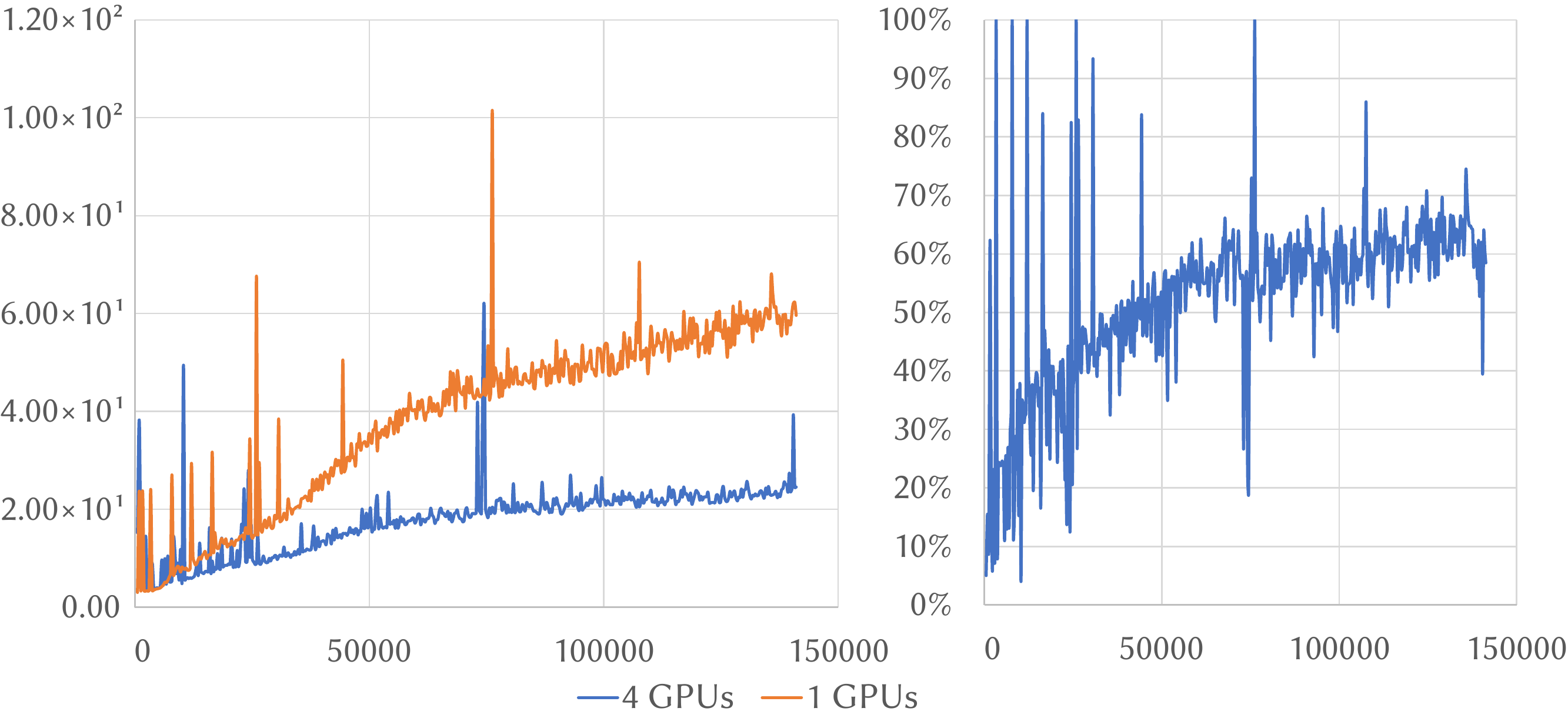}
    \caption{\textbf{Left:} Timing (in milliseconds) per frame of the Fountain with maximum $143.5$K particles, as the number of particles increases; \textbf{Right:} four-GPU Efficiency as the number of particles increases. Spikes are caused by memory reallocations.}
    \label{fig:fountain_realtime}
\end{figure}
\begin{figure}[htb]
\centering
    \includegraphics[width=1.0\linewidth]{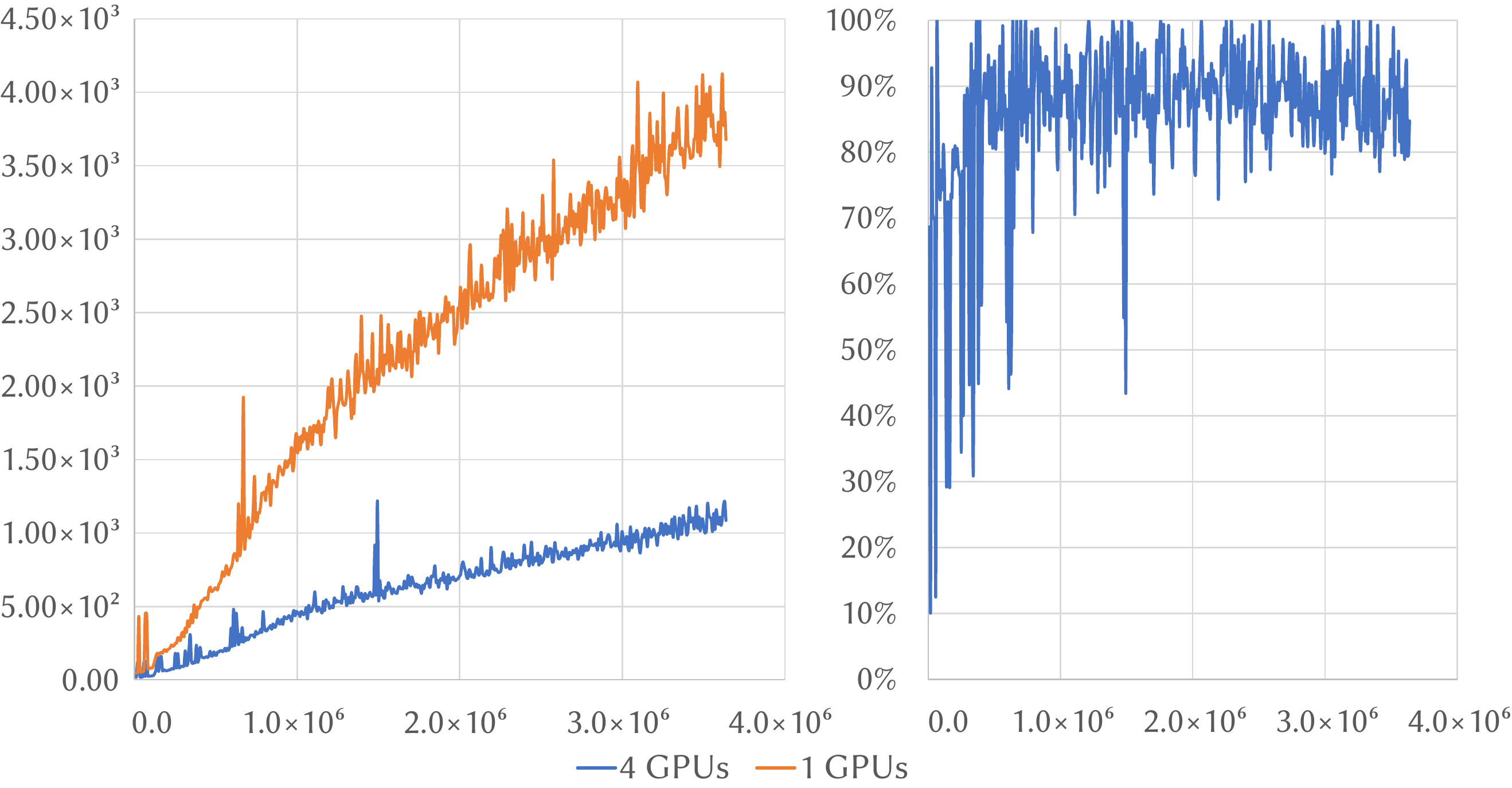}
    \caption{\textbf{Left:} Timing (in milliseconds) per frame of \emph{Fountain} with maximal 3.6M particles, as the number of particles increases; \textbf{Right:} four-GPU Efficiency as the number of particles increases. Spikes are caused by memory reallocations.}
    \label{fig:fountain_offline}
\end{figure}

Our pipeline can also be applied to scenarios where new particles are generated every frame, which can be difficult for pipelines adopting G2P2G. We simulate a fountain (\figref{fountain_rendering}) similar to the one in Fei et al.~\shortcite{fei2021revisiting}, and profile two different scales: one that can be simulated in real-time ($55.9$ frames per second on average with four GPUs) with maximal $143.5$K particles (\figref{fountain_realtime}), and another one that is simulated offline with maximal $3.6$M particles (\figref{fountain_offline}). Each simulated frame is $\frac{1}{60}$ second.

In both two cases, we adopt the ASFLIP integrator~\cite{fei2021revisiting} and a weakly-compressible material~\cite{tampubolon2017multi} with bulk modulus $1.0\times10^5 \text{dyne}/\text{cm}^2$. The ball-shaped water source is resampled at each frame with $27$ particles per cell. We adopt a cell size $0.66\text{cm}$ for the interactive case and $0.25\text{cm}$ for the offline case. The time-step sizes are automatically deducted~\cite{fang2018temporally} ($6.13\times10^{-4}$ on average for the interactive case; and $1.78\times10^{-4}$ on average for the offline case).

In this scenario, particles spray and travel rapidly, and thus the rebuild-mappings are performed more frequently ($11.77$ steps on average between two rebuild-mappings). As a result, the number of particles used to keep the simulation running at an interactive rate is relatively low. Even in this case, the four-GPU efficiency of our pipeline is around $52\%$. And it can further boost to $89\%$ when the number of particles increases to larger than one million.

\subsection{Crawling in Snow}
\begin{figure*}[t]
\centering
    \includegraphics[width=1.0\linewidth]{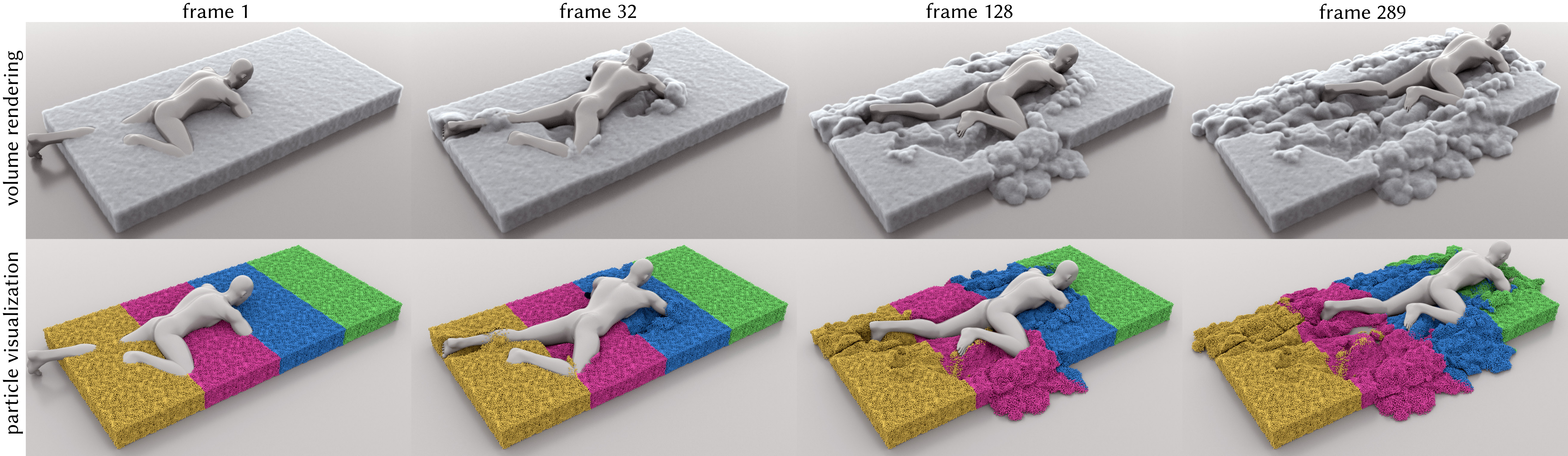}
    \caption{\textbf{Crawling in Snow.} A man crawls forward in the snow. This simulation has $1.33$M particles and can be simulated in real-time ($\geq 60$ frames per second) with four Tesla V100 GPUs. At the bottom, particles simulated with different GPUs are colored differently.}
    \label{fig:crawling_rendering}
\end{figure*}
We also profile our pipeline in a scenario with a more complicated animated boundary. We simulate a man crawling in the snow (\figref{crawling_rendering}). 
The animation of the man-shaped triangular mesh is converted to a sequence of levelsets during the precomputation. We model the snow as an NACC material~\cite{wolper2019cd}, with bulk modulus $\kappa=5.00\times10^5$, Poisson ratio $\nu=0.3$, friction coefficient $\text{M}=0.8$, cohesion coefficient $\beta=0.9$, and hardening coefficient $\xi=5.0$ (refer to Wolper et al.~\shortcite{wolper2019cd} for the notations). We adopt a cell size $1.35$cm. The time-step size is automatically deducted~\cite{fang2018temporally} ($5.46\times10^{-4}$ on average).

\begin{figure}[htb]
\centering
    \includegraphics[width=1.0\linewidth]{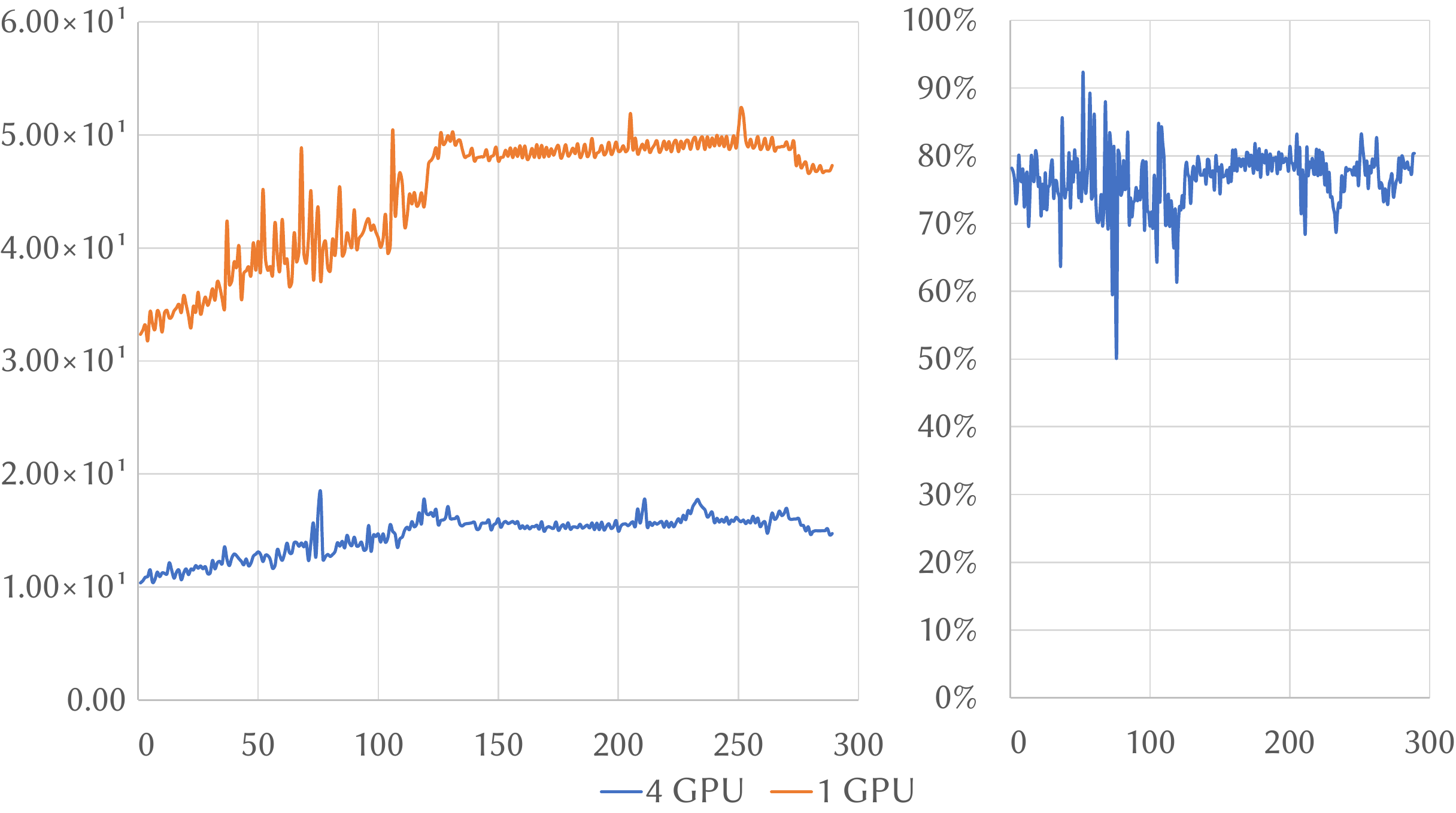}
    \caption{\textbf{Left:} Timing (in milliseconds) per frame of \emph{Crawling in Snow} with 1.33M particles over different frames; \textbf{Right:} four-GPU Efficiency over different frames.}
    \label{fig:snow_crawling}
\end{figure}
The snow dynamics are relatively slow, which allows rebuild-mappings to be performed less frequently ($84.36$ steps on average between two rebuild-mappings). As a result, the performance is much better than the \emph{Fountain} and it allows us to simulate $1.33$M particles in real-time, i.e., $\geq 60$ frames per second with each simulated frame being $\frac{1}{60}$ second (\figref{snow_crawling}). And our pipeline achieves a significant four-GPU efficiency (on average $77\%$).

\section{Limitation and Future Work}
We introduce principles for optimizing MPM on both single-GPU and multi-GPUs and demonstrate that by following these principles our new pipeline achieves a dramatic performance boost compared to the state-of-the-art. 
However, we believe there is still room for improvement in many aspects and we list some of them as follows.

\paragraph{Workload (re)-balancing} With multi-GPUs, we partition and distribute the particles evenly on each GPU during initialization. The partition pattern is pre-decided and its impact on the performance is not thoroughly studied.
Neither do we consider dynamic workload re-balancing during simulation.
When particles from different GPUs are randomly mixed, e.g., a paddle spinning in the sand may mix the sand from different GPUs, each physical block may be shared by all the GPUs and the inter-GPU communication would become the new bottleneck.
Dynamic re-partition of the particles across multiple cards is still an open problem.
Furthermore, transporting particles between GPUs during the simulation also brings additional costs.

\paragraph{Adaptive, hierarchical data structure} We only consider the simulation of MPM with a single level of the (sparse) background grid and all particles are of the same size.
Gao et al.~\shortcite{gao2017adaptive} present an MPM simulation with a multi-level grid that supports adaptive refining and coarsening of different regions.
It can be even more challenging to optimize on multiple GPUs for adaptive MPM.

\paragraph{Variants of MPM} We focus on the performance analysis and optimization for explicit integration of traditional MPM. 
Other forms or variants of MPM, e.g., semi-implicit MPM~\cite{klar2016drucker}, augmented MPM~\cite{stomakhin2014augmented}, IQ-MPM~\cite{fang2020iq}, and MPM with codimensional objects~\cite{fei2018multi,fei2021revisiting,guo2018material,jiang2017anisotropic} involve additional components that are not studied in this work. And we leave the analysis and/or optimization of them as future work.

\paragraph{Validation on different devices} We only test our pipeline on Tesla V100 GPUs which is based on NVIDIA Volta architecture. There are many other GPU architectures (e.g., NVIDIA Turing, NVIDIA Ampere, and AMD and Intel GPUs) that have different on-chip memory designs (e.g., different sizes of register files and caches) and instruction sets~\cite{jia2019dissecting}. Hence, our pipeline and principles may need to be tweaked and validated on GPUs with other architectures.


\bibliographystyle{ACM-Reference-Format}
{
\setlength{\topsep}{0ex}
\setlength{\itemsep}{0ex}
\setlength{\partopsep}{0ex}
\setlength{\parsep}{0ex}
\bibliography{ref}
}
\end{document}